\documentclass[showkeys,nofootinbib]{revtex4-2}

\usepackage[utf8]{inputenc} 
\usepackage[T1]{fontenc} 
\usepackage[english]{babel} 
\usepackage{xcolor} 
\usepackage{amsmath,amsfonts,amssymb,amsthm,braket} 
\usepackage{graphicx} 
\usepackage{tikz} 
\usepackage{multirow} 
\usepackage{fancyhdr} 
\usepackage[capitalise]{cleveref} 
\usepackage{enumitem} 
\usepackage{titlesec} 
\usepackage{lipsum}
\usepackage{listings} 

\theoremstyle{definition}

\theoremstyle{plain}

\usetikzlibrary{arrows.meta}
\definecolor{asparagus}{rgb}{0.53, 0.66, 0.42}
\definecolor{burntorange}{rgb}{0.8, 0.33, 0.0}
\definecolor{armygreen}{rgb}{0.29, 0.33, 0.13}
\definecolor{atomictangerine}{rgb}{1.0, 0.6, 0.4}
\definecolor{bluebell}{rgb}{0.64, 0.64, 0.82}
\definecolor{blue(pigment)}{rgb}{0.2, 0.2, 0.6}

\newcommand{\diff}{\ensuremath{\mathrm{d}}}
\newcommand{\bal}{\begin{aligned}}
\newcommand{\eal}{\end{aligned}}

\renewcommand{\bm}[1]{\boldsymbol{#1}}

\titleformat{\subsubsection}
   {\normalfont\fontsize{9pt}{11pt}\selectfont\bfseries}
   {\thesubsubsection.}
   {1em}
   {\centering}
\titleformat*{\subsubsection}{\normalsize\itshape}

\begin{document}

\title{Three-body Forces in Oscillator Bases Expansion}

\author{\surname{Cyrille} Chevalier}
\email[E-mail: ]{cyrille.chevalier@umons.ac.be}
\thanks{ORCiD: 0000-0002-4509-4309}

\author{\surname{Selma} Youcef Khodja}
\email[E-mail: ]{selma.youcef.khodja@ulb.be}

\affiliation{Service de Physique Nucl\'{e}aire et Subnucl\'{e}aire,
Universit\'{e} de Mons,
UMONS Research Institute for Complex Systems,
Place du Parc 20, 7000 Mons, Belgium}
\date{\today}

\begin{abstract}
   The oscillator bases expansion stands as an efficient approximation method for the time-independent Schrödinger equation. The method, originally formulated with one non-linear variational parameter, can be extended to incorporate two such parameters. It handles both non- and semi-relativistic kinematics with generic two-body interactions.  In the current work, focusing on systems of three identical bodies, the method is generalised to include the management of a given class of three-body forces. The computational cost of this generalisation proves to not exceed the one for two-body interactions. The accuracy of the generalisation is assessed by comparing with results from Lagrange mesh method and hyperspherical harmonic expansions. Extensions for systems of $N$ identical bodies and for systems of two identical particles and one distinct are also discussed.
\end{abstract}

\maketitle


\section{Introduction}
\label{sec:intro}

Resolving the time-independent Schrödinger equation stands as a standard stage across various disciplines in physics, spanning from hadronic and nuclear to atomic physics. To accomplish this task, a wide range of approximation methods exists, each possessing distinct strengths and limitations. Among these, a significant family relies on the Rayleigh-Ritz variational method and its extension to multiple trial wave-functions, known as MacDonald's theorem \cite{macd33}. This study focuses on the oscillator bases expansion (OBE) with different sizes that uses harmonic oscillator eigenstates with two non-linear parameters as trial functions \cite{silv20}. This method allows one to solve accurately a wide spectrum of Hamiltonians, encompassing both non-relativistic or semi-relativistic kinematics. Moreover, the OBE easily manages conditions on the symmetry, the angular momentum and the parity of the desired solution by selecting the trial states. These features make the OBE particularly efficient in hadronic physics where semi-relativistic kinematics, given angular momentum, and given parity are often used \cite{flec88,sema94,sema05,klem12,noh21}. A number of results presented here have already been obtained in several papers. They are nevertheless recalled here so that the work is complete and self-contained.

Focusing on three-body systems, this paper extends the method, originally formulated for two-body interactions, to encompass a specific class of three-body forces. The presence of many-body interactions arises in various contexts, such as the description of ultra-cold helium clusters in atomic physics \cite{gatt11}, of baryon spectra in hadron phenomenology \cite{buis22,desp92,pepi92,dmit01,ferr95} and within potentials utilised in nuclear physics \cite{ishi17}. When the true structure of these interactions is difficult to implement, they can be replaced by phenomenological forms, such as the ones encompassed in the present work. This makes the management of this type of interactions a subject of interest. Moreover, the endeavor performed in this paper to extend the OBE to accommodate three-body forces also establishes a link with the hyperspherical harmonics expansion (HHE). This connection is established through the demonstration that the trial states can be decomposed into a finite number of harmonic oscillator eigenstates written in hyperspherical coordinates \cite{das16,rayn70}.

The subsequent sections are organized as follows. Section \ref{sec:theory} elucidates the theoretical concepts related to the OBE. Subsection \ref{ssec:3BHO} provides a review of harmonic oscillator eigenstates and their essential properties, along with an explanation of their decomposition within hyperspherical coordinates. In Subsection \ref{ssec:Ham}, the class of Hamiltonians amendable to the OBE is introduced, while Subsections \ref{ssec:basis} and \ref{ssec:matEl}  delineate the incorporation of symmetries into the bases and provides formulas to evaluate matrix elements, respectively. Section \ref{sec:tests} examines the accuracy and the complexity of the OBE in presence of three-body forces. Section \ref{sec:gene} proposes a direct extension of the developed procedure to compute three-body matrix elements for systems of $N$ identical particles. Section \ref{sec:conclu} offers concluding remarks. Throughout, natural units are assumed ($\hbar=1$ and $c=1$) and considerations are confined to spatial degrees of freedom.


\section{Theory}
\label{sec:theory}

Let us consider a Hamiltonian denoted as $H$ for which the spectrum is sought. Let $\{\ket{\phi_i}\}_{i\in\{1,...,N\}}$ denotes a general finite and orthonormal set of trial states. MacDonald's theorem \cite{macd33} asserts that the $i$-th lowest eigenvalue of the Hamiltonian matrix provides an upper bound for the $i$-th lowest eigenvalue of $H$. The eigenvectors of this matrix also serve as approximations for the eigenvectors of $H$.

Methods based on this theorem are therefore divided in two primary steps : the computation of the matrix elements $\bra{\phi_i}H\ket{\phi_j}$ and the subsequent diagonalisation of the resulting matrix. To refine the precision of the upper bound, parameters are often introduced into the trial states. The procedure is repeated for different parameter values, yielding multiple upper bounds. The most constraining upper bound, that is the lowest one, is retained.

In the following, the trial states $\ket{\phi_i}$ are specified as eigenstates of the harmonic oscillator Hamiltonian, forming the so-called OBE. This article is dedicated to the development of an efficient procedure for computing the matrix elements for this choice of trial states, particularly for systems of three particles and in the presence of three-body forces.


\subsection{The Harmonic Oscillator Eigenstates}
\label{ssec:3BHO}

Before to dive into the evaluation of matrix elements, the expression of the harmonic oscillator eigenstates as well as some of their properties have to be introduced. The harmonic oscillator Hamiltonian in a generic system of two three-dimensional coordinates, $\bm{x}$ and $\bm{y}$, is given by
\begin{equation}
    H_{\text{oh}} = \frac{\bm{p}^2}{2} + \frac{\bm{q}^2}{2} + \frac{\bm{x}^2}{2} + \frac{\bm{y}^2}{2}
    \label{eq::OHH}.
\end{equation}
Here, $\bm{p}$ and $\bm{q}$ represent the conjugate variables to $\bm{x}$ and $\bm{y}$ respectively. The coordinates are supposed to be dimensionless. In a first approach, this Hamiltonian can be analytically solved using the method of separation of variables. The resulting eigenfunctions, denoted as $\Phi^L_{n_xl_xn_yl_y}(\bm{x},\bm{y})$, and eigenvalues, denoted as $E^L_{n_xl_xn_yl_y}$, are given by
\begin{equation}
    \Phi^L_{n_xl_xn_yl_y}(\bm{x},\bm{y}) = \left[\phi_{n_xl_x}(\bm{x})\phi_{n_yl_y}(\bm{y})\right]_L \text{ and } E^L_{n_xl_xn_yl_y} = 2n_x+l_x+2n_y+l_y+3
    \label{eq::eig_2OH}
\end{equation}
The square brackets $[...]_L$ indicate that a total orbital angular momentum $L$ is provided to the state,
\begin{equation}
    \left[\phi_{n_xl_x}(\bm{x})\phi_{n_yl_y}(\bm{y})\right]_L = \sum_{m_xm_y} \braket{l_xm_xl_ym_y|LM} \phi_{n_xl_xm_x}(\bm{x})\phi_{n_yl_ym_y}(\bm{y})
    \label{eq::Clebsh-Gordan}
\end{equation}
where the symbols $\braket{l_xm_xl_ym_y|LM}$ represent Clebsh-Gordan coefficients \cite{vars88}. The total angular momentum projection $M$ is omitted in the left-hand side notation because none of the matrix elements considered in the following will depend on this quantum number\footnote{As these are summed during the coupling, the $m_x$ and $m_y$ quantum numbers are also excluded from the left-hand side notation.}. Following the standard rules for angular momentum coupling, the $l_x$, $l_y$ and $L$ quantum numbers must satisfy $|l_x-l_y|\leq L \leq l_x+l_y$. The functions $\phi_{nlm}(\bm{r})$ corresponds to the well-known one-body harmonic oscillator eigenfunctions which are defined as
\begin{equation}
    \phi_{nlm}(\bm{r})=R_{nl}(r)Y_{lm}(\hat{r}) \text{ with } R_{nl}(r)=\sqrt{\frac{2(n!)}{\Gamma(n+l+3/2)}}\, r^l e^{-\frac{r^2}{2}} L^{l+1/2}_n(r^2).
    \label{eq::eig_1OH}
\end{equation}
Here $r$ and $\hat{r}$ represent the modulus and the angles of the vector $\bm{r}$, $Y_{lm}$ stands for the standard spherical harmonics \cite{vars88} and $L^{l+1/2}_n$ represents the generalized Laguerre polynomials \cite{abra64}. This expression allows to interpret $n_x$ and $l_x$ (resp. $n_y$ and $l_y$) as the radial and orbital quantum numbers related to the $\bm{x}$ (resp. $\bm{y}$) coordinate. In the following, because this combination will regularly reappear, $2n_x+l_x+2n_y+l_y$ is named the number of quanta of $\Phi$, denoted $Q$. 

The $\Phi$ wave functions are the ones used as trial functions in the oscillator bases expansion for three-body systems. Efficiently evaluating matrix elements such as 
$\bra{\Phi^L_{n_x'l_x'n_y'l_y'}(\bm{x},\bm{y})}\mathcal{O}(\bm{x},\bm{y},\bm{p},\bm{q})\ket{\Phi^L_{n_xl_xn_yl_y}(\bm{x},\bm{y})}$ is therefore a primary concern for this method. The rest of this section is devoted to some properties of the harmonic oscillator eigenstates that will provide support in this regard.

\subsubsection{Talmi's integral technique}

In general, it is possible to reduce the aforementioned matrix elements to the evaluation of matrix elements on one-body harmonic oscillator eigenfunctions \eqref{eq::eig_1OH},
\begin{equation}
    \bra{\phi_{n'l'm'}(\bm{r})}\mathcal{O}(ar)\ket{\phi_{nlm}(\bm{r})} = \delta_{l'l} \delta_{m'm} \int r^2\diff r\ R_{n'l}(r)\,\mathcal{O}(ar)\,R_{nl}(r). 
    \label{eq::matEl_10H}
\end{equation}
with $a$ a real parameter which has been made explicit for further use. The remaining integral can be evaluated using the Talmi's integral technique \cite{brod60},
\begin{equation}
    \int r^2\diff r\ R_{n'l}(r)\,\mathcal{O}(ar)\,R_{nl}(r) = \sum_{p\,=\,\frac{l+l'}{2}}^{n+n'+\frac{l+l'}{2}} B(n'l,nl,p)\, I_p(\mathcal{O},a)
    \label{eq::TalmiIntegral1}
\end{equation}
with
\begin{equation}
     I_p(\mathcal{O},a) = \frac{2}{\Gamma(p+3/2)}\int \diff r\, r^{2p+2} e^{-r^2} \mathcal{O}(ar).
    \label{eq::TalmiIntegral2}
\end{equation}
A closed formula for the $B(n'l;nl;p)$ coefficient is provided in \cite{brod60}. Since these coefficients depend only on the quantum numbers within the set of trial states, they can be computed once, stored and  subsequently retrieved whenever needed. For many functions $\mathcal{O}$ (including cases of physical interest), the integrals $I_p(\mathcal{O},a)$ admit analytical expressions, eliminating the need for numerical integration (see \ref{sec::TalmiInt}). In all cases, the use of formulas \labelcref{eq::TalmiIntegral1,eq::TalmiIntegral2} enables rapid and accurate evaluations of the matrix elements \eqref{eq::matEl_10H}.

\subsubsection{Momentum matrix element}

The $\mathcal{O}$ function in equation \eqref{eq::matEl_10H} depend solely on a position variable. Consequently, the direct application of Talmi's integral technique is not feasible when dealing with a momentum variable. However, as the Fourier transform of a one-body harmonic oscillator eigenfunctions remains a one-body harmonic oscillator eigenfunctions (up to a phase factor),
\begin{equation}
    \frac{1}{(2\pi)^{3/2}}\int \diff^3r\, e^{-i\bm{p}\bm{r}} \phi_{nlm}(\bm{r}) = (-i)^{2n+l} \phi_{nlm}(\bm{p}),
    \label{eq::FT_1HO}
\end{equation}
this technique can still be applied by switching to momentum representation. For quadratic functions of momentum, which are particularly relevant as they are associated with non-relativistic kinetic energy, the following relation can also be employed to simplify calculations \cite[relation 3.10]{mosh69},
\begin{equation}
    \bal
    \bra{\phi_{n'l'm'}(\bm{r})}\bm{p}^2\ket{\phi_{nlm}(\bm{r})} =  \delta_{l'l}\delta_{m'm}\Big((2n+l+3/2)&\delta_{n'n} \\
    + \sqrt{n(n+l+1/2)}&\delta_{n'+1\,n} + \sqrt{n'(n'+l+1/2)}\delta_{n'n+1}\Big).
    \label{eq::p2_1HO}
    \eal
\end{equation}

\subsubsection{Inversion and exchange of coordinates}

Thanks to the properties of spherical harmonics \cite[section 5.5.2]{vars88}, one can show that an inversion of the $\bm{x}$ and $\bm{y}$ coordinates in the $\Phi^L_{n_xl_xn_yl_y}(\bm{x},\bm{y})$ functions simply produces additional real phase factors,
\begin{equation}
    \Phi^L_{n_xl_xn_yl_y}(-\bm{x},-\bm{y}) = (-1)^{l_x+l_y} \Phi^L_{n_xl_xn_yl_y}(\bm{x},\bm{y}).
    \label{eq::eigHO_inv_xy}
\end{equation}
The $\bm{x}$ and $\bm{y}$ coordinates can also be exchanged. Using symmetry properties of Clebsh-Gordan coefficients \cite[section 8.4.3]{vars88}, one can show that
\begin{equation}
    \Phi^L_{n_xl_xn_yl_y}(\bm{y},\bm{x}) = (-1)^{L-l_x-l_y} \Phi^L_{n_yl_yn_xl_x}(\bm{x},\bm{y}).
    \label{eq::eigHO_ex_xy}
\end{equation}

\subsubsection{Rotation of the coordinates}

Instead of directly solving the Hamiltonian \eqref{eq::OHH}, one may try beforehand a change of coordinate. This Hamiltonian is symmetrical under the following coordinate transformation,
\begin{equation}
    \begin{cases}
    \bm{\tilde{x}} = \cos\beta\,\bm{x} + \sin\beta\,\bm{y}\\
    \bm{\tilde{y}} = -\sin\beta\,\bm{x} + \cos\beta\,\bm{y}
    \end{cases}.
    \label{eq::rotCoor}
\end{equation}
As a consequence, all the functions $\Phi^L_{n_xl_xn_yl_y}(\bm{\tilde{x}},\bm{\tilde{y}})$ are also eigenfunctions of \eqref{eq::OHH} with $2n_x+l_x+2n_y+l_y+3$ as eigenvalue. This ensures that any $\Phi^L_{n_xl_xn_yl_y}(\bm{\tilde{x}},\bm{\tilde{y}})$ can be expressed as a linear combination of $\Phi^L_{n_x'l_x'n_y'l_y'}(\bm{x},\bm{y})$ functions that share the same energy eigenvalue than $\Phi^L_{n_xl_xn_yl_y}(\bm{\tilde{x}},\bm{\tilde{y}})$,
\begin{equation}
    \Phi^L_{n_xl_xn_yl_y}(\bm{\tilde{x}},\tilde{\bm{y}}) = \sum_{n_x'l_x'n_y'l_y'} \braket{n_x'l_x'n_y'l_y';L|n_xl_xn_yl_y;L}_{\beta} \Phi^L_{n_x'l_x'n_y'l_y'}(\bm{x},\bm{y})
    \label{eq::BMexp}
\end{equation}
where the summations on $n_x'$, $l_y'$, $n_y'$ and $l_y'$ are restricted to terms with a number of quanta equal to $2n_x+l_x+2n_y+l_y$. This constraint ensures that the summation is finite. The coefficients denoted as $\braket{n_x'l_x'n_y'l_y';L|n_xl_xn_yl_y;L}_{\beta}$ in equation \eqref{eq::BMexp} are known as Brody-Moshinsky coefficients with angle $\beta$. These can be computed recursively using the formula proposed in \cite{silv85}.

\subsubsection{Passage to Hyperspherical coordinates}

The Hamiltonian \eqref{eq::OHH} can also be solved in hyperspherical coordinates \cite{rayn70}. The two three-dimensional $\bm{x}$ and $\bm{y}$ vectors are replaced by a single six-dimensional vector for which hyperspherical coordinates are used \cite{das16,rayn70}. An hyperradius is defined
\begin{equation}
    \rho^2 = \bm{x}^2 + \bm{y}^2,
     \label{eq::rho}
\end{equation}
as well as five hyperangles. Four among the five angles are simply chosen as the usual polar and azimuthal angles of the $\bm{x}$ and $\bm{y}$ vectors. These will therefore respectively be denoted $\theta_x$, $\theta_y$, $\varphi_x$ and $\varphi_y$. The fifth angle, denoted $\alpha$, is defined as follows\footnote{This definition differs from the one used in \cite{rayn70}. To compare our relations with the ones from this reference, $x$ and $y$ have to be exchanged.},
\begin{align}
    & x = \rho \sin\alpha, & & y =\rho \cos\alpha.
     \label{eq::alpha}
\end{align}
Because both $x$ and $y$ are positive definite, $\alpha$ lies in between $0$ and $\pi/2$. For notation convenience, in the following, the set of angles $(\theta_x,\theta_y,\varphi_x,\varphi_y,\alpha)$ will be denoted as $\Omega$. Switching to this system of coordinates, the harmonic oscillator Hamiltonian \eqref{eq::OHH} in position representation reads
\begin{equation}
    H_{\text{oh}} = -\frac{1}{2}\left(\frac{\partial^2}{\partial \rho^2} + \frac{5}{\rho}\frac{\partial}{\partial \rho}-\frac{1}{\rho^2}\Delta_{\Omega}\right) +\frac{\rho^2}{2}.
     \label{eq::HO_hypersph}
\end{equation}
Above $\Delta_{\Omega}$ is the hyperspherical Laplacian operator in six dimensions. The full expression of this operator is depicted in \cite{das16,rayn70}. Its eigenfunctions, known as hyperspherical harmonics, reads
\begin{equation}
\bal
    \mathcal{Y}_{K}^{l_xl_yLM}&(\Omega) = \sum_{m_xm_y} \braket{l_xm_xl_ym_y|LM} N_K^{l_xl_y} \sin^{l_x}\alpha\, \cos^{l_y}\alpha\, P^{\left(l_x+\frac{1}{2},l_y+\frac{1}{2}\right)}_{n}(\cos 2\alpha) Y_{l_xm_x}(\theta_x,\varphi_x) Y_{l_ym_y}(\theta_y,\varphi_y)\\
    \text{ with } &N_K^{l_xl_y}=\left(\frac{2n!(K+2)(n+l_x+l_y+1)!}{\Gamma(n+l_x+3/2)\Gamma(n+l_y+3/2)}\right)^{1/2} \text{ and } n=\frac{K-l_x-l_y}{2}\label{eq::hypersph}
\eal 
\end{equation}
with the corresponding eigenvalue $K(K+4)$. Above, $P^{(a,b)}_n$ denotes a Jacobi polynomial \cite{abra64}. The expression \eqref{eq::HO_hypersph} is the one of a one-body harmonic oscillator Hamiltonian in a six-dimensional space. Separating the hyperradial and hyperangular parts of the equation, the following eigenfunctions are obtained \cite{rayn70}, 
\begin{equation}
    \Psi_{NK}^{l_xl_yL}(\rho,\Omega) = \mathcal{R}_{NK}(\rho)\, \mathcal{Y}_{K}^{l_xl_yLM}(\Omega) \text{ with }\mathcal{R}_{NK}(\rho) = \sqrt{\frac{2(N!)}{\Gamma(K+N+3)}} \,e^{-\frac{\rho^2}{2}} \rho^K L_N^{K+2}(\rho^2),
    \label{eq::hyperEig1}
\end{equation}
the associated eigenvalues being $    E_{NK}^{l_xl_ym_xm_y} = 2N+K+3.$
By construction, states $\psi_{NK}^{l_xl_ym_xm_y}$ have a total orbital angular momentum $L$ as well as individual orbital angular momenta $l_x$ and $l_y$.

Both set of functions $\Psi_{NK}^{l_xl_yL}(\rho,\Omega)$ and $\Phi^L_{n_xl_xn_yl_y}(\bm{x},\bm{y})$ are eigenstates of the harmonic oscillator Hamiltonian, of the total orbital angular momentum operator and of the two individual angular momenta operators. The $\Phi^L_{n_xl_xn_yl_y}(\bm{x},\bm{y})$ functions can therefore be written as a linear combination of all the $\Psi_{NK}^{l_xl_yL}(\rho,\Omega)$ functions with energy eigenvalue $E^L_{n_xl_xn_yl_y}$,
\begin{equation}
    \Phi^L_{n_xl_xn_yl_y}(\bm{x},\bm{y}) = \sum_{N\,K} \braket{n_xn_y|NK}_{l_xl_y} \Psi_{NK}^{l_xl_yL}(\rho,\Omega)
    \label{eq::hyperExp}
\end{equation}
where the summations on $N$ and $K$ are restricted to terms where $2N+K=2n_x+l_x+2n_y+l_y$. The coefficients $\braket{n_xn_y|NK}_{l_xl_y}$ are defined by the following overlap integral,
\begin{equation}
\braket{n_xn_y|NK}_{l_xl_y} = \int \diff^3x\,\diff^3y\,\Psi_{NK}^{l_xl_yL\, ^*}(\rho,\Omega)\Phi^{L}_{n_xl_xn_yl_y}(\bm{x},\bm{y}).
\label{eq::hyperCoef_def}
\end{equation}
In the following, these coefficients will be referred as hyperspherical coefficients. A closed formula to compute them is demonstrated in \ref{sec::hyperCoef}.


\subsection{Specification of The Hamiltonian}
\label{ssec:Ham}

In this section, the harmonic oscillator eigenstates are utilized to solve the following three-body Hamiltonian,
\begin{equation}
    H = T(\bm{p}_1,\bm{p}_2,\bm{p}_3) + V_{12}(r_{12}) +  V_{13}(r_{13}) +  V_{23}(r_{23}) + W\!\left(\sqrt{r_{12}^2 + r_{13}^2 + r_{23}^2}\right) \text{ with } r_{ij}=|\bm{r}_i - \bm{r}_j|.
    \label{eq::H}
\end{equation}
The variable $\bm{r}_i$ and $\bm{p}_i$ denote the position and momentum of the $i$-th particle in the system, respectively. General $V_{ij}$ functions are employed to introduce two-body interactions in the system, and the $W$ function is used to handle a certain class of three-body forces. Although not in the most general form, the implemented three-body forces have been successfully used to describe physical systems like baryons \cite{ferr95,dmit01,pepi92,buis22} and helium clusters \cite{gatt11}. For the kinetic part of $H$, both non-relativistic and semirelativistic kinematics can be considered. In the non-relativistic case, the kinetic energy is given by
\begin{equation}
    T_{\text{nr}}(\bm{p}_1,\bm{p}_2,\bm{p}_3) = \frac{\bm{p}_1^2}{2m_1} + \frac{\bm{p}_2^2}{2m_2} + \frac{\bm{p}_3^2}{2m_3} - \frac{\bm{P}^2}{2M} \text{ with }\bm{P} = \bm{p}_1+\bm{p}_2+\bm{p}_3 \text{ and }M=m_1+m_2+m_3.
    \label{eq::T_nr}
\end{equation}
Here, the fourth term explicitly accounts for the removal of the center of mass contribution. In the semi-relativistic case, the expression for the kinetic energy is
\begin{equation}
    T_{\text{r}}(\bm{p}_1,\bm{p}_2,\bm{p}_3) = \sqrt{\bm{p}_1^2 + m_1^2} + \sqrt{\bm{p}_2^2 + m_2^2} + \sqrt{\bm{p}_3^2 + m_3^2}.
    \label{eq::T_r}
\end{equation}
In this case, the center of mass motion has not been removed. To avoid any disturbance to the energy levels caused by center of mass excitation, the Hamiltonian is considered to be evaluated in the system's rest frame by manually setting $\bm{P}=\bm{0}$.

To develop an oscillator bases expansion able to solve the Hamiltonian \eqref{eq::H}, dimensionless Jacobi coordinates that are scaled with two variational parameters denoted $a$ and $b$ are chosen \cite{silv20,nunb77}. These Jacobi coordinates are defined as
\begin{align}
    &a\bm{x} = \bm{r}_1-\bm{r}_2, & &b\bm{y} = \frac{m_1\bm{r}_1+m_2\bm{r}_2}{m_{12}} - \bm{r}_3, & &\bm{R}=\frac{m_1\bm{r}_1+m_2\bm{r}_2+m_3\bm{r}_3}{M},
\end{align}
where $m_{12}$ stands for $m_1+m_2$. The momenta conjugate to $\bm{x}$, $\bm{y}$ and $\bm{R}$ are respectively given by
\begin{align}
    &\frac{\bm{p}}{a}=\frac{m_2\bm{p}_1 - m_1\bm{p}_2}{m_{12}}, & &\frac{\bm{q}}{b}=\frac{m_3(\bm{p}_1+\bm{p}_2) - m_{12}\bm{p}_3}{M}, & &\bm{P}=\bm{p}_1 + \bm{p}_2 + \bm{p}_3.
\end{align}
The Hamiltonian can explicitly be written in the new system of coordinates. The non-relativistic kinetic energies becomes
\begin{equation}
    T_{\text{nr}}(\bm{p},\bm{q}) = \frac{\bm{p}^2}{2\mu_p} + \frac{\bm{q}^2}
    {2\mu_q} \text{ with } \mu_p = \frac{a^2m_1m_2}{m_{12}} \text{ and }\mu_q = \frac{b^2m_3m_{12}}{M},
    \label{eq::T_nr_2}
\end{equation}
the semi-relativistic one becomes
\begin{equation}
    T_{\text{r}}(\bm{p},\bm{q}) = \sqrt{\left(\frac{m_1}{m_{12}}\frac{\bm{q}}{b}+\frac{\bm{p}}{a}\right)^2+m_1^2} + \sqrt{\left(\frac{m_2}{m_{12}}\frac{\bm{q}}{b}-\frac{\bm{p}}{a}\right)^2+m_2^2} + \sqrt{\frac{\bm{q}^2}{b^2}+m_3^2},
    \label{eq::T_r_2}
\end{equation}
and finally the arguments of both two-body and three-body interactions are modified,
\begin{equation}   
    V_{12}(a\bm{x}) + V_{13}\left(-b\bm{y}-\frac{m_2}{m_{12}}a\bm{x}\right) + V_{23}\left(b\bm{y}-\frac{m_1}{m_{12}}a\bm{x}\right),
\end{equation}  
\begin{equation}
    W\!\left(\sqrt{a^2\left(1+\frac{m_1^2+m_2^2}{m_{12}^2}\right)\bm{x}^2 + 2b^2\bm{y}^2 + 2ab\left(\frac{m_2-m_1}{m_{12}}\right)\bm{x}\cdot\bm{y}}\right).
    \label{eq::W}
\end{equation}


\subsection{Setup of the basis}
\label{ssec:basis}

As the center of mass motion has been removed in $H$, both the variables $\bm{R}$ and $\bm{P}$ do not enter in any parts of the Hamiltonian, thereby reducing the number of variables to two. The remaining $\bm{x}$ and $\bm{y}$ coordinates can be used as a pair of three-dimensional coordinates on which harmonic oscillator eigenstates from Section \ref{ssec:3BHO} can depend. In the current Section, these functions will be used to built a set of trial states that fits with the features of the system under consideration. 
\vspace{2mm}

\subsubsection{Orbital angular momentum and parity}

First, in most problems, a solution with a given total orbital angular momentum is expected. To ensure that the provided approximation satisfies this requirement, the set of $\Phi$ functions is restricted to the ones whose $L$ quantum number is in agreement with the expected total orbital angular momentum. Similarly, for some applications, a parity eigenstate is also expected. As the parity transformation, denoted $\Pi$, inverts the positions of each particles, its action on Jacobi coordinates is given by $\Pi\, \bm{x}=-\bm{x}$ and $\Pi\, \bm{y}=-\bm{y}$. Therefore, the action of the parity operator on the trial functions can directly be deduced from relation \eqref{eq::eigHO_inv_xy},
\begin{equation}
    \Pi\,\Phi^L_{n_xl_xn_yl_y}(\bm{x},\bm{y}) = \Phi^L_{n_xl_xn_yl_y}(-\bm{x},-\bm{y}) = (-1)^{l_x+l_y}\Phi^L_{n_xl_xn_yl_y}(\bm{x},\bm{y}).
\end{equation}
The $\Phi$ functions are already parity eigenstates and the corresponding eigenvalue is given by the parity of $l_x + l_y$. Filling the set of trial states with even (resp. odd) sum of $l_x$ and $l_y$ ensure that the provided approximation will be a positive (resp. negative) parity eigenstate.
\vspace{2mm}

\subsubsection{Symmetry under exchange of particles}

In presence of identical particles in the system, symmetries under the exchange of pairs of particles have to be introduced.  The state of a system containing two identical particles must be either symmetric (for bosons) or anti-symmetric (for fermions) under their permutation. Let us choose particle $1$ and $2$ as the two identical particle and denote $\mathbb{P}_{12}$ the operator that encodes their permutation. 
Acting on the Jacobi coordinates, this operator reverses $\bm{x}$ and lets $\bm{y}$ unchanged, 
\begin{align}
    &\mathbb{P}_{12}\, \bm{x} = -\bm{x}, & &\mathbb{P}_{12}\, \bm{y} = \bm{y}.
    \label{eq::perm_jacobi}
\end{align}
As a consequence, the action of $\mathbb{P}_{12}$ on a $\Phi$ function can directly be obtained from property \eqref{eq::eigHO_inv_xy},
\begin{equation}
    \mathbb{P}_{12} \Phi^L_{n_xl_xn_yl_y}(\bm{x},\bm{y}) = \Phi^L_{n_xl_xn_yl_y}(-\bm{x},\bm{y}) = (-1)^{l_x} \Phi^L_{n_xl_xn_yl_y}(\bm{x},\bm{y}).
\end{equation}
A $\Phi$ function is already an eigenstate of $\mathbb{P}_{12}$ and the corresponding eigenvalue is given by the parity of its $l_x$ quantum number. In conclusion, in presence of two identical particles, the set of trial states must be filled with $\Phi$ functions of same $l_x$ parity (odd in presence of fermions and even in presence of bosons).

In presence of three identical particles, the situation is a bit more complicated. The state of the system must now be completely (anti)symmetric under any exchange of particles. Hopefully, because any permutation of three particles can be expressed in terms of two transpositions, it is sufficient to (anti)symmetrise on the exchange of particles $1$ and $2$ and of particles $2$ and $3$. For the first one, the task has already been done in the previous paragraph: a correct $\mathbb{P}_{12}$ symmetry is ensured by considering only states of a given $l_x$ parity. The rest of the calculations considers this selection as carried out. The (anti)symmetrisation must now be performed on the exchange of particles $2$ and $3$, denoted $\mathbb{P}_{23}$. The action of this operator on Jacobi coordinates is given by
\begin{align}
    &\mathbb{P}_{23}\,\bm{x} = \frac{\bm{x}}{2} + \frac{b}{a}\bm{y}, & &\mathbb{P}_{23}\,\bm{y} = \frac{3a}{4b}\bm{x} - \frac{\bm{y}}{2}.\label{eq::P23}
\end{align}
With such a coordinate transformation, the $\Phi$ functions will not be eigenstates of this operator. To obtain (anti)symmetric functions, one has to compute the matrix elements of the $\mathbb{P}_{23}$ operator and to subsequently diagonalise it,
\begin{equation}
    \bal
     \bra{\Phi^{L'}_{n_x'l_x'n_y'l_y'}(\bm{x},\bm{y})}\mathbb{P}_{23}\ket{\Phi^L_{n_xl_xn_yl_y}(\bm{x},\bm{y})} =\ &\braket{\Phi^{L'}_{n_x'l_x'n_y'l_y'}(\bm{x},\bm{y})|\Phi^L_{n_xl_xn_yl_y}(\bm{x'},\bm{y'})}\\ &\hspace{2.5cm}\text{ with } \bm{x'}= \frac{\bm{x}}{2} + \frac{b}{a}\bm{y} \text{ and } \bm{y'}= \frac{3a}{4b}\bm{x} - \frac{\bm{y}}{2}.
     \eal
\end{equation}
It has been shown in \cite{silv20} that the eigenstates of $\mathbb{P}_{23}$ contain a finite number of $\Phi$ states only if $b=\sqrt{3}a/2$. Assuming that, \eqref{eq::P23} becomes a rotation and the $\mathbb{P}_{23}$ matrix elements can be evaluated using property \eqref{eq::BMexp},
\begin{equation}
    \bra{\Phi^{L'}_{n_x'l_x'n_y'l_y'}(\bm{x},\bm{y})}\mathbb{P}_{23}\ket{\Phi^L_{n_xl_xn_yl_y}(\bm{x},\bm{y})} = \delta_{L'L} (-1)^{l_x'+l_y'+L} \braket{n_y'l_y'n_x'l_x';L'|n_xl_xn_yl_y;L}_{\pi/6}.
    \label{eq::P13_matEl}
\end{equation}
By definition, the Brody-Moshinsky coefficient in this relation is zero if both number of quanta differs. As announced, $\mathbb{P}_{23}$ mixes together a finite number of states, the ones with the same total orbital angular momentum, the same parity and the same number of quanta. Therefore, to exactly get symmetrized states, it is sufficient to built the $\mathbb{P}_{23}$ matrix up to an arbitrary number of quanta and to diagonalise it. The set of trial states can then be filled with all the obtained eigenvectors whose eigenvalue corresponds to the expected symmetry (namely $+1$ for three-boson systems and $-1$ for three-fermion systems). As a consequence, for three identical particles, the set $\Phi$ functions cannot be truncated randomly: all the $\Phi$ functions until a given number of quanta are necessarily involved in the calculations.

\subsubsection{Convenient truncation for the basis}

Except for systems of three identical particles, it seems that the oscillator basis can be truncated everywhere. However, because property \eqref{eq::BMexp} will be used during the evaluation of matrix elements, a given $\Phi$ function will be decomposed as a linear combination of all the $\Phi$ functions that have the same number of quanta. Therefore, it is more convenient to include from the beginning all these functions in the calculations. So, even for systems of three different particles, the recommended sets of trial states contain all the $\Phi$ functions until a given number of quanta, denoted $Q_{\text{max}}$. The higher $Q_{\text{max}}$ is, the more accurate the approximation produced by the set is but the larger the Hamiltonian matrix is.


\subsection{Calculation of Matrix Elements}
\label{ssec:matEl}

The set of trial states being chosen, the matrix elements of the Hamiltonian have now to be evaluated. This goal will be completed thanks to the properties enumerated in section \ref{ssec:3BHO}. Kinetic energies, two-body interactions and three-body interactions are considered separately.

\subsubsection{Non-relativistic kinetic energy}

Non-relativistic kinetics \eqref{eq::T_nr_2} are considered first. The following matrix elements have to be evaluated,
\begin{equation}
    \bal
    \bra{\Phi^{L'}_{n_x'l_x'n_y'l_y'}(\bm{x},\bm{y})} \left(\frac{\bm{p}^2}{2\mu_p} + \frac{\bm{q}^2}
    {2\mu_q} \right) \ket{\Phi^{L}_{n_xl_xn_yl_y}(\bm{x},\bm{y})} = \frac{1}{2\mu_p}&\bra{\Phi^{L'}_{n_x'l_x'n_y'l_y'}(\bm{x},\bm{y})} \bm{p}^2 \ket{\Phi^{L}_{n_xl_xn_yl_y}(\bm{x},\bm{y})} \\ 
    &\hspace{1cm} + \frac{1}{2\mu_q}\bra{\Phi^{L'}_{n_x'l_x'n_y'l_y'}(\bm{x},\bm{y})} \bm{q}^2 \ket{\Phi^{L}_{n_xl_xn_yl_y}(\bm{x},\bm{y})}.
    \eal \label{eq::Tnr_matEl}
\end{equation}
As both terms can be evaluated similarly, let us focus on the first one. To start with, $\Phi$ functions are made explicit using relations \eqref{eq::eig_2OH} and \eqref{eq::Clebsh-Gordan}. Because the $\bm{p}^2$ operator only acts on the $\bm{x}$ part of the $\phi$ function, the orthonormalisation of the $\phi$ function can be used to get rid of the $\bm{y}$ dependence.
In addition, considering that $\bm{p}^2$ is a scalar operator, Clebsh-Gordan coefficients can be removed using results from \cite[section 8.1.1]{vars88},
\begin{equation}
    \bra{\Phi^{L'}_{n_x'l_x'n_y'l_y'}(\bm{x},\bm{y})} \bm{p}^2 \ket{\Phi^{L}_{n_xl_xn_yl_y}(\bm{x},\bm{y})} = \delta_{L'L} \delta_{n_y'n_y}\delta_{l_y'l_y}\delta_{l_x'l_x}\bra{\phi_{n_x'l_x0}(\bm{x})}\bm{p}^2\ket{\phi_{n_xl_x0}(\bm{x})}. \label{eq::matEl_p2}
\end{equation}
Relation \eqref{eq::p2_1HO} can then be used to evaluate the residual matrix element. Collating with the analogous result for the $\bm{q}^2$ term, the following analytical expression for the kinetic energy matrix energy is obtained,
\begin{equation}
    \bra{\Phi^{L'}_{n_x'l_x'n_y'l_y'}(\bm{x},\bm{y})} \left(\frac{\bm{p}^2}{2\mu_p} + \frac{\bm{q}^2}
    {2\mu_q} \right) \ket{\Phi^{L}_{n_xl_xn_yl_y}(\bm{x},\bm{y})} = \delta_{L'L}\delta_{l_x'l_x}\delta_{l_y'l_y}\left(\frac{\delta_{n_y'n_y}}{a^2} \left(K_{\bm{p}}\right)_{n_x'n_x} + \frac{\delta_{n_x'n_x}}{b^2}\left(K_{\bm{q}}\right)_{n_y'n_y}\right) 
\end{equation}
with
\begin{equation}
    \bal
    & \left(K_{\bm{p}}\right)_{n_x'n_x} = \frac{m_{12}}{2m_1m_2} \left( (2n_x+l_x+3/2)\delta_{n_x'n_x}
    + \sqrt{n_x(n_x+l_x+1/2)}\delta_{n_x'+1\,n_x} + \sqrt{n_x'(n_x'+l_x+1/2)}\delta_{n_x'n_x+1}\right),\\
    & \left(K_{\bm{q}}\right)_{n_y'n_y} = \frac{M}{2m_{12}m_3} \left( (2n_y+l_y+3/2)\delta_{n_y'n_y}
    + \sqrt{n_y(n_y+l_y+1/2)}\delta_{n_y'+1\,n_y} + \sqrt{n_y'(n_y'+l_y+1/2)}\delta_{n_y'n_y+1}\right).
    \eal
\end{equation}
The $K_{\bm{p}}$ and $K_{\bm{q}}$ matrices being independent of the $a$ and $b$ variationnal parameters, they can be computed once for all at the begining of the optimisation. Let us also mention that if the three particles are identical, due to symmetry properties of the trial states, both terms from \eqref{eq::Tnr_matEl} proves to be equal, thereby simplifying calculations.

\subsubsection{Two-body potential matrix elements}

The two-body potential divides in three terms, one for each pair of particles. The one related to particle $1$ and $2$ only depends on the $\bm{x}$ coordinate. Using the same arguments than for relation \eqref{eq::matEl_p2} in the treatment of the non-relativistic kinetic energy, the $\bm{y}$ coordinate and the Clebsh-Gordan coefficients can be eliminated
\begin{equation}
    \bal
    \bra{\Phi^{L'}_{n_x'l_x'n_y'l_y'}(\bm{x},\bm{y})} V_{12}(a\bm{x}) \ket{\Phi^{L}_{n_xl_xn_yl_y}(\bm{x},\bm{y})} = \delta_{L'L} \delta_{n_y'n_y}\delta_{l_y'l_y}
    \bra{\phi_{n_x'l_x'0}(\bm{x})} V_{12}(a\bm{x})\ket{\phi_{n_xl_x0}(\bm{x})}. \label{eq::matEl_12}
    \eal
\end{equation}
The remaining matrix element can be evaluated using the Talmi's integral technology developed in section \ref{ssec:3BHO}. 

Both other two-body interactions depend explicitly on the $\bm{x}$ and $\bm{y}$ coordinates, thereby making their evaluation less straightforward. Next developments focus on the $V_{23}$ matrix elements,
\begin{equation}
    \bra{\Phi^{L'}_{n_x'l_x'n_y'l_y'}(\bm{x},\bm{y})}V_{23}\left(b\bm{y}-\frac{m_1}{m_{12}}a\bm{x}\right)\ket{\Phi^{L}_{n_xl_xn_yl_y}(\bm{x},\bm{y})}.
    \label{eq::matEl_23pot}
\end{equation}
To evaluate these matrix elements, the $\bm{x}$ and $\bm{y}$ coordinates can be rotated as in relation \eqref{eq::rotCoor}. An angle $\beta_1$ is defined so that
\begin{equation}
    \bal
    &\cos(\beta_1) = \frac{m_{12}}{\gamma_1}b, && &\sin(\beta_1) = \frac{m_{1}}{\gamma_1}a.
    \eal
\end{equation}
with $\gamma_1 = \sqrt{b^2m_{12}^2+a^2m_1^2}$. New $\bm{\tilde x}$ and $\bm{\tilde y}$ coordinates are obtained by rotating $-\bm{x}$ and $\bm{y}$ with angle $-\beta_1$,
\begin{equation}
    \bal
    \bm{\tilde{x}} = \frac{m_{12}}{\gamma_1}\left(-b\bm{x} - \frac{m_{1}}{m_{12}}a\bm{y}\right),\\
    \bm{\tilde{y}} = \frac{m_{12}}{\gamma_1}\left(-\frac{m_{1}}{m_{12}}a\bm{x} + b\bm{y}\right).
    \eal
\end{equation} The angle and the rotation have been chosen so that the argument of the $V_{23}$ function is colinear with the $\bm{\tilde y}$ coordinate. The inverse change of coordinates rotates $\bm{\tilde x}$ and $\bm{\tilde y}$ with angle $\beta_1$ to recover $-\bm{x}$ and $\bm{y}$. Using relations \eqref{eq::eigHO_inv_xy} and  \eqref{eq::BMexp}, the $\Phi$ functions can be rewritten in terms of $\bm{\tilde x}$ and $\bm{\tilde y}$,
\begin{equation}
    \Phi^{L}_{n_xl_xn_yl_y}(\bm{x},\bm{y}) = (-1)^{l_x}\sum_{\nu_x\lambda_x\nu_y\lambda_y} \braket{\nu_x\lambda_x\nu_y\lambda_y;L|n_xl_xn_yl_y;L}_{\beta_1} \Phi^{L}_{\nu_x\lambda_x\nu_y\lambda_y}(\bm{\tilde x},\bm{\tilde y}).
    \label{eq::matEl_BM}
\end{equation}
When inserted in \eqref{eq::matEl_23pot}, equation \eqref{eq::matEl_BM} gives
\begin{equation}
    \bal
    &\bra{\Phi^{L'}_{n_x'l_x'n_y'l_y'}(\bm{x},\bm{y})}V_{23}\left(b\bm{y}-\frac{m_1}{m_{12}}a\bm{x}\right)\ket{\Phi^{L}_{n_xl_xn_yl_y}(\bm{x},\bm{y})}\\
    &\hspace{3cm} = (-1)^{l_x+l_x'} \sum_{\nu_x\lambda_x\nu_y\lambda_y}\sum_{\nu_x'\lambda_x'\nu_y'\lambda_y'} \braket{\nu_x'\lambda_x'\nu_y'\lambda_y';L'|n_x'l_x'n_y'l_y';L'}_{\beta_1}\braket{\nu_x\lambda_x\nu_y\lambda_y;L|n_xl_xn_yl_y;L}_{\beta_1}\\ &\hspace{8.5cm}    \bra{\Phi^{L'}_{\nu_x'\lambda_x'\nu_y'\lambda_y'}(\bm{\tilde x},\bm{\tilde y})}V_{23}\left(\frac{\gamma_1}{m_{12}}\,\bm{\tilde y}\right)\ket{\Phi^{L}_{\nu_x\lambda_x\nu_y\lambda_y}(\bm{\tilde x},\bm{\tilde y})}
    \eal
\end{equation}
which can be simplified as
\begin{equation}
    \bal
    &\bra{\Phi^{L'}_{n_x'l_x'n_y'l_y'}(\bm{x},\bm{y})}V_{23}\left(b\bm{y}-\frac{m_1}{m_{12}}a\bm{x}\right)\ket{\Phi^{L}_{n_xl_xn_yl_y}(\bm{x},\bm{y})}\\
    &\hspace{3cm} = \delta_{L'L} (-1)^{l_x+l_x'} \sum_{\nu_x\lambda_x\nu_y\lambda_y\nu_y'\lambda_y'} \braket{\nu_x\lambda_x\nu_y'\lambda_y';L'|n_x'l_x'n_y'l_y';L'}_{\beta_1}\braket{\nu_x\lambda_x\nu_y\lambda_y;L|n_xl_xn_yl_y;L}_{\beta_1}\\ &\hspace{10.5cm}    \bra{\phi_{\nu_y'\lambda_y'0}(\bm{\tilde y})}V_{23}\left(\frac{\gamma_1}{m_{12}}\,\bm{\tilde y}\right)\ket{\phi_{\nu_y\lambda_y0}(\bm{\tilde y})}.
    \label{eq::matEl_V23}
    \eal
\end{equation}
This formula, together with the Talmi's integral technique, allows for an efficient evaluation of $V_{23}$ matrix elements. However, because $\beta_1$ explicitly depends on the $a$ and $b$ variational parameters, Brody-Moshinsky coefficients will have to be recomputed at each step of the optimisation.

The $V_{13}$ matrix element can be evaluated using the same technique than for $V_{23}$. First, a $\beta_2$ angle is introduced so that 
\begin{equation}
    \bal
    &\cos(\beta_2) = \frac{m_{12}}{\gamma_2}b, && &\sin(\beta_2) = \frac{m_2}{\gamma_2}a.
    \eal
\end{equation}
with $\gamma_2 = \sqrt{b^2m_{12}^2+a^2m_2^2}$. New $\bm{\tilde x}$ and $\bm{\tilde y}$ coordinates are defined by rotating $-\bm{y}$ and $\bm{x}$ with angle $-\beta_2$. The $\bm{\tilde x}$ coordinate appears to be proportional to the argument of $V_{13}$ while relations \eqref{eq::eigHO_inv_xy}, \eqref{eq::eigHO_ex_xy} and \eqref{eq::BMexp} are used to rewrite the states in terms of the new coordinates. Finally, simplifying the remaining matrix elements, a relation similar to \eqref{eq::matEl_V23} is obtained\footnote{The formula proposed in \cite{silv20} differs a bit from the one presented here. The equivalence of the two formula can be shown using symmetry properties of Brody-Moshinsky coefficients.},
\begin{equation}
    \bal
    &\bra{\Phi^{L'}_{n_x'l_x'n_y'l_y'}(\bm{x},\bm{y})}V_{13}\left(-b\bm{y}-\frac{m_2}{m_{12}}a\bm{x}\right)\ket{\Phi^{L}_{n_xl_xn_yl_y}(\bm{x},\bm{y})}\\
    &\hspace{3cm} = \delta_{L'L}(-1)^{l_x+l_x'} \sum_{\nu_x\lambda_x\nu_y\lambda_y\nu_x'\lambda_x'} \braket{\nu_x'\lambda_x'\nu_y\lambda_y;L'|n_y'l_y'n_x'l_x';L'}_{\beta_2}\braket{\nu_x\lambda_x\nu_y\lambda_y;L|n_yl_yn_xl_x;L}_{\beta_2}\\ &\hspace{10.5cm}    \bra{\phi_{\nu_x'\lambda_x'0}(\bm{\tilde x})}V_{13}\left(\frac{\gamma_2}{m_{12}}\,\bm{\tilde x}\right)\ket{\phi_{\nu_x\lambda_x0}(\bm{\tilde x})}.
    \label{eq::matEl_V13}
    \eal
\end{equation}
Again, $\beta_2$ depends on $a$ and $b$, forcing to recompute Brody-Moshinsky coefficients at each step of the optimisation. However, in presence of at least two identical particles (chosen as $1$ and $2$), the $V_{23}$ and $V_{13}$ matrix elements are shown to be equal. Therefore, the evaluation of one of these two matrix elements and the included calculation of Brody-Moshinsky coefficient can be avoided. The same accounts for three identical particles: the three two-body potential matrix elements are shown to be equal and the single calculation of $V_{12}$ is sufficient.

\subsubsection{Semi-relativistic matrix elements}
\label{sssec::RelKin}

The semi-relativistic matrix elements can be computed by switching to momentum representation. Making use of relation \eqref{eq::FT_1HO}, the evaluation of these matrix elements reduces to
\begin{equation}
    \bal
    \bra{\Phi^{L'}_{n_x'l_x'n_y'l_y'}(\bm{x},\bm{y})}T_{r}(\bm{p},\bm{q})\ket{\Phi^{L}_{n_xl_xn_yl_y}(\bm{x},\bm{y})} = i&^{2n_x'+l_x'+2n_y'+l_y'}(-i)^{2n_x+l_x+2n_y+l_y}\\
    &\hspace{0.5cm}\int \diff^3p\, \diff^3q \left(\Phi^{L'}_{n_x'l_x'n_y'l_y'}(\bm{p},\bm{q})\right)^*T_{r}(\bm{p},\bm{q})\,\Phi^{L}_{n_xl_xn_yl_y}(\bm{p},\bm{q}).
    \eal
\end{equation}
The integral in the right-hand side is extremely similar to the ones evaluated for two-body matrix elements. The $T_{r}$ function divides also in three terms. The third one only depends on the $\bm{q}$ variable and can then be evaluated similarly to the $V_{12}$ matrix element,
\begin{equation}
    \bal
    i^{2n_x'+l_x'+2n_y'+l_y'}(-i)^{2n_x+l_x+2n_y+l_y} &\int \diff^3p\, \diff^3q \left(\Phi^{L'}_{n_x'l_x'n_y'l_y'}(\bm{p},\bm{q})\right)^*\left(\frac{\bm{q}^2}{b^2}+m_3^2\right)^{1/2}\Phi^{L}_{n_xl_xn_yl_y}(\bm{p},\bm{q}) \\
    = (i)&^{2n_y'+l_y'}(-i)^{2n_y+l_y} \delta_{L'L} \delta_{n_x'n_x}\delta_{l_x'l_x}
    \int \diff^3q\,\left(\phi_{n_y'l_y'0}(\bm{q})\right)^* \left(\frac{\bm{q}^2}{b^2}+m_3^2\right)^{1/2}\phi_{n_yl_y0}(\bm{q}).
    \eal \label{eq::matEl_TR1}
\end{equation}
The residual integral has the structure that Talmi's integral technique is able to evaluate (see \ref{sec::TalmiInt} for an analytical formula). The two other terms can be evaluated following the same procedure than for the $V_{23}$ and $V_{13}$ two-body matrix elements. Let us consider the following term at first,
\begin{equation}
    \int \diff^3p\, \diff^3q \left(\Phi^{L'}_{n_x'l_x'n_y'l_y'}(\bm{p},\bm{q})\right)^*\left(\left(\frac{m_1}{m_{12}}\frac{\bm{q}}{b}+\frac{\bm{p}}{a}\right)^2+m_1^2\right)^{1/2}\Phi^{L}_{n_xl_xn_yl_y}(\bm{p},\bm{q}).
\end{equation}
New $\bm{\tilde p}$ and $\bm{\tilde q}$ coordinates are defined by rotating $\bm{p}$ and $-\bm{q}$ with angle $-\beta_1$. On the one hand, the $\bm{\tilde p}$ coordinate turns out to be proportional to the linear combination of $\bm{p}$ and $\bm{q}$ that occurs in the matrix element,
\begin{equation}
    \bm{\tilde p} = \eta_1\left(\frac{m_1}{m_{12}}\frac{\bm{q}}{b}+\frac{\bm{p}}{a}\right) \text{ with }\eta_1 = \frac{abm_{12}}{\sqrt{b^2m_{12}^2+a^2m_1^2}}.
\end{equation}
On the other hand the $\Phi$ functions can be rewritten in terms of the new coordinates making use of relations
\eqref{eq::eigHO_inv_xy} and \eqref{eq::BMexp}. Combining these results, eliminating the Clebsh-Gordan coefficients and proceeding to integration on $\bm{\tilde q}$, the aforementioned matrix element can be evaluated by computing\footnote{\label{note}The result in \cite{silv20} slightly differs from the one presented here because, in this reference, parity conservation has been used to simplify the phases.}
\begin{equation}
    \bal
    \int \diff^3p\, \diff^3q &\left(\Phi^{L'}_{n_x'l_x'n_y'l_y'}(\bm{p},\bm{q})\right)^*\left(\left(\frac{m_1}{m_{12}}\frac{\bm{q}}{b}+\frac{\bm{p}}{a}\right)^2+m_1^2\right)^{1/2}\Phi^{L}_{n_xl_xn_yl_y}(\bm{p},\bm{q})\\ 
    &\hspace{1cm}= (-1)^{l_y+l_y'}\sum_{\nu_x'\lambda_x'\nu_y'\lambda_y'}\sum_{\nu_x\lambda_x\nu_y\lambda_y} \braket{\nu_x'\lambda_x'\nu_y'\lambda_y';L'|n_x'l_x'n_y'l_y';L'}_{\beta_1}\braket{\nu_x\lambda_x\nu_y\lambda_y;L|n_xl_xn_yl_y;L}_{\beta_1}\\
    &\hspace{5cm} \delta_{\nu_y'\nu_y}\delta_{\lambda_y'\lambda_y}\int \diff^3\tilde p \left(\phi_{\nu_x'\lambda_x'0}(\bm{\tilde p})\right)^*\left(\frac{\bm{\tilde p}^2}{\eta_1^2}+m_1^2\right)^{1/2}\phi_{\nu_x\lambda_x0}(\bm{\tilde p}).
    \eal\label{eq::matEl_TR2}
\end{equation}
Once again the remaining integral can be evaluated using Talmi's integral technology. Finally, same tricks are used to compute the third matrix element,
\begin{equation}
    \int \diff^3p\, \diff^3q \left(\Phi^{L'}_{n_x'l_x'n_y'l_y'}(\bm{p},\bm{q})\right)^*\left(\left(\frac{m_2}{m_{12}}\frac{\bm{q}}{b}-\frac{\bm{p}}{a}\right)^2+m_2^2\right)^{1/2}\Phi^{L}_{n_xl_xn_yl_y}(\bm{p},\bm{q}).
\end{equation}
New coordinates are defined by rotating $\bm{p}$ and $\bm{q}$ with angle $-\beta_2$ and, using the same arguments, the six-dimensional integral is reduced to\footref{note}
\begin{equation}
    \bal
    \int \diff^3p\, \diff^3q &\left(\Phi^{L'}_{n_x'l_x'n_y'l_y'}(\bm{p},\bm{q})\right)^*\left(\left(\frac{m_2}{m_{12}}\frac{\bm{q}}{b}-\frac{\bm{p}}{a}\right)^2+m_2^2\right)^{1/2}\Phi^{L}_{n_xl_xn_yl_y}(\bm{p},\bm{q})\\ 
    &\hspace{1cm}= \sum_{\nu_x'\lambda_x'\nu_y'\lambda_y'}\sum_{\nu_x\lambda_x\nu_y\lambda_y} \braket{\nu_x'\lambda_x'\nu_y'\lambda_y';L'|n_x'l_x'n_y'l_y';L'}_{\beta_2}\braket{\nu_x\lambda_x\nu_y\lambda_y;L|n_xl_xn_yl_y;L}_{\beta_2}\\
    &\hspace{5cm} \delta_{\nu_y'\nu_y}\delta_{\lambda_y'\lambda_y}\int \diff^3\tilde p \left(\phi_{\nu_x'\lambda_x'0}(\bm{\tilde p})\right)^*\left(\frac{\bm{\tilde p}^2}{\eta_2^2}+m_2^2\right)^{1/2}\phi_{\nu_x\lambda_x0}(\bm{\tilde p})
    \eal \label{eq::matEl_TR3}
\end{equation}
with $\eta_2 = abm_{12}/\sqrt{b^2m_{12}^2+a^2m_2^2}$ and where the remaining integrals can still be evaluated with Talmi's technique. In presence of two identical particles, thanks to symmetry properties of the states, this matrix element proves to be equal to the previous one, allowing to avoid its evaluation. In presence of three identical particles, the three matrix elements proves to be equal and only formula \eqref{eq::matEl_TR1} is needed.

\subsubsection{Three-body potential matrix elements}
\label{sec::3BMatEl}

Due to the $\bm x \cdot \bm y$ term in \eqref{eq::W}, the three-body potential matrix elements seem harder to evaluate than the two-body ones. However in presence of at least two identical particles (chosen as $1$ and $2$), this term disappears thanks to the vanishing $m_2-m_1$ coefficient,
\begin{equation}
    W\!\left(\sqrt{a^2\left(1+\frac{m_1^2+m_2^2}{m_{12}^2}\right)\bm{x}^2 + 2b^2\bm{y}^2 + 2ab\left(\frac{m_2-m_1}{m_{12}}\right)\bm{x}\cdot\bm{y}}\right) = W\!\left(\sqrt{2b^2\bm{y}^2 + 3/2a^2\bm{x}^2}\right) .
\end{equation}
Under this assumption, the matrix elements can be evaluated naively, using directly the expressions \eqref{eq::eig_2OH} to \eqref{eq::eig_1OH} for the $\Phi$ functions,
\begin{equation}
    \bal
    \bra{\Phi^{L'}_{n_x'l_x'n_y'l_y'}(\bm{x},\bm{y})}&W\!\left(2b^2\bm{y}^2 + 3/2a^2\bm{x}^2\right)\ket{\Phi^{L}_{n_xl_xn_yl_y}(\bm{x},\bm{y})} \\ &= \delta_{l_x'l_x}\delta_{l_y'l_y}\int x^2\diff x \, y^2\diff y\, R_{n_x'l_x'}(x)R_{n_y'l_y'}(y) W\!\left(\sqrt{2b^2y^2 + 3/2a^2 x^2}\right) R_{n_xl_x}(x)R_{n_yl_y}(y).
    \eal \label{eq::dbl_int}
\end{equation}
Properties of Clebsh-Gordan coefficients \cite[section 8.1.1]{vars88} and spherical harmonics orthonormality relation \cite[sections 5.1.4]{vars88} have been used to get rid of the angular dependence. Within this approach, matrix elements are given by two-dimensional integrals, resulting in a considerably higher numerical cost than for the evaluation of matrix elements for two-body potentials. 

However, if in addition the constraint on $a$ and $b$ for three identical particles is imposed, the argument of $W$ proves to be proportional to the hyperradius defined in \eqref{eq::rho},
\begin{equation}
    W\!\left(\sqrt{2b^2\bm{y}^2 + 3/2a^2\bm{x}^2}\right)= W\!\left(\sqrt{3/2}a\rho\right) .
\end{equation}
It is worth noting that this relation requires that $m_1=m_2$ and $b=\sqrt{3}a/2$ but not that $m_2=m_3$. Therefore, the following developments can also be used for systems of two identical particles provided that the second non-linear variational parameter is gave up. The matrix elements of a function that only depends on the hyperradius can easily be evaluated using property \eqref{eq::hyperExp},
\begin{equation}
\bal
    &\bra{\Phi^{L'}_{n_x'l_x'n_y'l_y'}(\bm{x},\bm{y})}W\!\left(\sqrt{3/2}a\rho\right)\ket{\Phi^{L}_{n_xl_xn_yl_y}(\bm{x},\bm{y})} \\
    & \hspace{1cm} = \sum_{N'K'NK} \braket{n_x'n_y'|N'K'}_{l_x'l_y'} \braket{n_xn_y|NK}_{l_xl_y}\bra{\Psi^{l_x'l_y'L'}_{N'K'}(\rho,\Omega)}W\!\left(\sqrt{3/2}a\rho\right)\ket{\Psi^{l_xl_yL}_{NK}(\rho,\Omega)}.
\eal \label{eq::matEl_hyperCoef}
\end{equation}
Expressions of $\Psi$ functions are given in \eqref{eq::hyperEig1}. In the remaining matrix elements, five of the six integrals are simplified using the orthogonality of hyperspherical harmonics \cite[relation (3.24)]{das16} and Clebsh-Gordan properties \cite[section 8.1.1]{vars88}. The remaining integrals reads (volume element in hyperspherical coordinates is given in \eqref{eq::hyper_volEl})
\begin{equation}
    \bal
    \bra{\Psi^{l_x'l_y'L'}_{N'K'}(\rho,\Omega)}W\!\left(\sqrt{3/2}a\rho\right)\ket{\Psi^{l_xl_yL}_{NK}(\rho,\Omega)} = \delta_{L'L} \delta_{l_x'l_x} \delta_{l_y'l_y} \delta_{K'K} \int \diff \rho\, \rho^5 \mathcal{R}_{N'K}(\rho)W\!\left(\sqrt{3/2}a\rho\right) \mathcal{R}_{NK}(\rho).
    \eal \label{eq::matEl_hyperCoord}
\end{equation}
This relation is extremely similar to relation \eqref{eq::matEl_10H} and can be evaluated using the same strategy, resulting in a modified Talmi's integral technique. The proof of the formula for $B(nl;n'l';p)$ coefficients given in \cite{brod60} is based on an explicit expression of Laguerre polynomials which encompasses both integers and half-integers indices, factorials having simply to be replaced by gamma functions \cite{abra64}. Therefore, even if the half-integer $l+1/2$ index in the Laguerre polynomial is replaced by an integer $K+2$ index, this formula remains valid.
It might have required an adaptation because normalisation coefficients from \eqref{eq::eig_1OH} and from \eqref{eq::hyperEig1} are different but these prove to cancel each other in the modified demonstration. As a consequence, the following formula is obtained for the integral from \eqref{eq::matEl_hyperCoord},
\begin{equation}
\bal
    &\int \diff \rho\, \rho^5 \mathcal{R}_{N'K}(\rho)W\!\left(\sqrt{3/2}a\rho\right) \mathcal{R}_{NK}(\rho) = \sum_{p = \frac{2K+3}{2}}^{N' +N+\frac{2K+3}{2}}  B(N'\,K+3/2; N\, K+3/2;p) I_p(W,\sqrt{3/2}a)
\eal \label{eq::HyperTalmi}
\end{equation}
with $I_p$ defined in \eqref{eq::TalmiIntegral2}. Together, formulas \eqref{eq::matEl_hyperCoef}, \eqref{eq::matEl_hyperCoord} and \eqref{eq::HyperTalmi} allow for an efficient evaluation of three-body potential matrix elements. It as been numerically checked with different $W$ that these ones provide the same results than relation \eqref{eq::dbl_int}.


\section{Tests of the Method}
\label{sec:tests}

The basis states having been chosen and the formulas for the matrix elements having been obtained, the method is ready to be implemented. This section is devoted to various tests aimed at assessing the accuracy of the method. These tests will not be performed on systems for which the OBE is a particularly efficient method, such as systems with semi-relativistic kinematics. On the contrary, they are performed on systems for which other approximation methods, namely the hyperspherical harmonic expansion (HHE) and the Lagrange mesh method (LMM), can supply a very accurate point of comparison. Reviews of these two methods can respectively be found in \cite{kiev08,marc20} and in \cite{baye15}. These tests also aim to encompass a diverse range of Hamiltonians that includes three-body forces. Comparisons are made with data from existing literature. The computational complexity of the OBE in presence of three-body forces is also discussed at the end of the section.

In the following, a condensed notation $\ket{\sigma;n;L^P}$ will be used to denote the eigenstates of the tested Hamiltonians. The label $\sigma$ denotes the symmetry of the state ($+1$ for a symmetric state, $-1$ for an anti-symmetric state), $P$ denotes its parity ($+$ for an even state, $-$ for an odd state) and $L$ denotes its total angular momentum. Lastly, the non-zero integer $n$ serves to differentiate and order the states depending on their energy.


\subsection{Accuracy Tests}
\label{ssec:ATest}

At first, tests can be performed on systems only bounded with three-body forces. A system of three non-relativistic bosons interacting through an attractive three-body Gaussian potential is considered,
\begin{align}
    &V_{12}(r) = V_{23}(r) = V_{13}(r) = 0, && W(\rho) = -3^{4/3}\,e^{-\rho^2/27}.\label{eq::first_acc_test}
\end{align}
Unit masses are used. Parameters have been chosen to allow the comparison with results obtained in \cite[p.87]{baye15} using LMM in hyperspherical coordinates. Only the ground state energy is supplied in the text but equations can easily be implemented to produce excited energies too. The ground state energies given by the OBE for this system and for different sizes of the basis are compiled in Table \ref{tab::gauss_att_gr}. Energies for the three lowest states from the OBE are compared to the ones from LMM in Table \ref{tab::gauss_comp}. The OBE is able to reproduce the energy spectrum up to nine to six digits depending on the energy level. Let us mention that the method used in \cite{baye15} is particularly efficient for such a system whose potential only depends on the hyperradius variable \eqref{eq::rho}. Hyperspherical equations become uncoupled and the system is entirely described by a single-variable hyperradial equation which is solved with the LMM.
\begin{table}
    \centering
    \begin{tabular}{ l l l l}
    $Q_{\text{max}}$\hspace{0.2cm} &$N$\hspace{1cm} & $a$ \hspace{0.7cm} & $E_{gs}$ \\
    \hline \hline
    $6$ & $7$ & $1.6695$ & $-1.739\,737\,863$ \\
    $8$ & $11$ & $1.6921$ & $-1.739\,828\,778$ \\
    $10$ & $16$ & $1.6868$ & $-1.739\,828\,773$ \\
    $12$ & $23$ & $1.6365$ & $-1.739\,830\,590$ \\
    $14$ & $31$ & $1.6365$ & $-1.739\,830\,808$ \\
    $16$ & $41$ & $1.6365$ & $-1.739\,830\,913$ \\
    $18$ & $53$ & $1.6365$ & $-1.739\,830\,929$ \\
    $20$ & $67$ & $1.6365$ & $-1.739\,830\,936$ \\
    $22$ & $83$ & $1.6365$ & $-1.739\,830\,937$ \\
    $24$ & $102$ & $1.6365$ & $-1.739\,830\,938$ \\
    \hline
    \end{tabular}
    \caption{Ground state energies, $E_{gs}$, for a system of three identical bosons interacting with potentials \eqref{eq::first_acc_test} are given depending on the basis size, $N$. For $Q_{\text{max}}=6,8,10,12$, optimisation on the non-linear parameter $a$ has been performed for $Q=Q_{\text{max}}$. For $Q_{\text{max}}>12$, the $a$ value is computed for $Q=12$. The ground state energy provided in \cite{baye15} is $1.739\,830\,938$.}
    \label{tab::gauss_att_gr}
\end{table}
\begin{table}
    \centering
    \begin{tabular}{l|c c}
     & \hspace{1mm} OBE & LMM \\
    \hline \hline
    $\ket{1;1;0^+}$ & \hspace{1mm} $-1.739\,830\,938$ \hspace{1mm} & $-1.739\,830\,938$\\
    $\ket{1;2;0^+}$ & \hspace{1mm}   $-0.552\,311\,353$ \hspace{1mm} & $-0.552\,311\,965$\\
    $\ket{1;1;2^+}$ & \hspace{1mm}  $-0.373\,040\,428$ \hspace{1mm} & $-0.373\,040\,920$
    \end{tabular}
    \vspace{2mm}
    \caption{Energies for the three lowest states for a system of three identical bosons interacting with potentials \eqref{eq::first_acc_test}. Results from the OBE and from LMM are compared. The maximal number of quanta used in the basis is $Q_{\text{max}}=24$ with $a$ computed for $Q=12$.}
    \label{tab::gauss_comp}
\end{table}

In addition to Gaussian interactions, three-body power potentials can easily be implemented too. A system of three non-relativistic bosons interacting through a three-body Colombian potential is considered,
\begin{align}
    &V_{12}(r) = V_{23}(r) = V_{13}(r) = 0, && W(\rho) = -\frac{3}{\rho}.\label{eq::sec_acc_test}
\end{align}
Unit masses are also used. This system is chosen to investigate the effect of a divergent potential rather than for its physical relevance. Accurate eigenenergies can again be obtained by solving equations from \cite{baye15}. These are compared to results from the OBE in Table \ref{tab::coul_comp}. Degeneracies between $\ket{1;3;0^+}$ and $\ket{1;1;2^+}$ as well as  between $\ket{1;1;1^-}$ and $\ket{1;1;3^-}$ are observed in the spectrum. The first is inherent to the Coulombian shape but the second occurs for any three-body potential and can be explained by the pure dependence of the system on the hyperradius. The aforementioned hyperradial equation which rules the dynamics of the system only depends on the $K$ quantum number of the hyperspherical harmonic, thereby producing degeneracies in the spectrum. Concerning accuracy, the OBE seems less efficient than in the previous test. For two-body potential, it is already well-known than divergences deteriorate the accuracy of the OBE. This feature seems to apply for three-body forces too. More accurate results can of course be obtained by increasing the number of quanta.
\begin{table}
    \centering
    \begin{tabular}{l|c c}
     & \hspace{1mm} OBE & LMM \\
    \hline \hline
    $\ket{1;1;0^+}$ & \hspace{1mm} $-0.239\,912\,74$ & $-0.239\,999\,98$\\
    $\ket{1;2;0^+}$ & \hspace{1mm} $-0.121\,949\,51$ & $-0.122\,448\,97$\\
    $\ket{1;1;2^+}$ & \hspace{1mm} $-0.074\,067\,53$ & $-0.074\,074\,07$\\
    $\ket{1;3;0^+}$ & \hspace{1mm} $-0.072\,931\,73$ & $-0.074\,074\,07$\\
    $\ket{1;1;1^-}$ & \hspace{1mm} $-0.049\,584\,24$ & $-0.049\,586\,78$\\
    $\ket{1;1;3^-}$ & \hspace{1mm} $-0.049\,584\,24$ & $-0.049\,586\,78$
    \end{tabular}
    \caption{Energies in arbitrary units of the six lowest states for a system of three identical bosons interacting with potentials \eqref{eq::sec_acc_test}. Results from the OBE and from HHE are compared. The maximal number of quanta used in basis is $Q_{\text{max}}=28$ with $a$ computed for $Q=16$.}
    \label{tab::coul_comp}
\end{table}

More physical systems, including both three- and two-body interactions can also be investigated. A first test is performed on a system of three non-relativistic identical bosons interacting with two-body attractive Coulomb interactions and confined by a three-body linear potential,
\begin{align}
    &V_{12}(r) = V_{23}(r) = V_{13}(r) = - \frac{1}{r}, && W(\rho) = \frac{1}{2} \rho.\label{eq::third_acc_test}
\end{align}
A more sophisticated version of this potential is used for the description of baryonic bound states in \cite{caps86}. Unit masses are used.
The comparison is conducted with results from an HHE \cite{PC}. Energies and mean-values of the inter-particle distances\footnote{Due to symmetry property, this mean-value is independent of the chosen pair of particles.} are given for the three lowest states in Table \ref{tab::coul_lin_comp}. As already mentioned, for very divergent potentials, such as the Coulombic potential, the OBE often exhibits reduced accuracy. In the current test, the method proved to be able to provide energies and mean-values with three significant digits. 
\begin{table}
    \centering
    \begin{tabular}{l|l l}
    $\ket{1;1;0^+}$ & \hspace{1mm} OBE \hspace{3mm} & HHE \cite{PC} \\
    \hline \hline
    $E$ & \hspace{1mm} $0.363$ & $0.360$ \\
    $\braket{r}$ & \hspace{1mm}  $1.368$ & $1.367$
    \end{tabular}\hfill
    \begin{tabular}{l|l l}
    $\ket{1;2;0^+}$ & \hspace{1mm} OBE \hspace{3mm} & HHE \cite{PC} \\
    \hline \hline
    $E$ \hspace{1mm} & \hspace{1mm} $1.953$ & $1.947$ \\
    $\braket{r}$ & \hspace{1mm} $2.220$ & $2.216$
    \end{tabular}\hfill
    \begin{tabular}{l|l l}
    $\ket{1;1;2^+}$ & \hspace{1mm} OBE \hspace{3mm} & HHE \cite{PC} \\
    \hline \hline
    $E$ & \hspace{1mm} $2.397$ & $2.395$ \\
    $\braket{r}$ & \hspace{1mm} $2.368$ & $2.366$
    \end{tabular}
    \caption{Energies and mean values in arbitrary units for the inter-particles distance of the three lowest states for a system of three identical bosons interacting with potentials \eqref{eq::third_acc_test}. Results from the OBE and from HHE are compared. The maximal number of quanta used in basis is $Q_{\text{max}}=24$ with $a$ computed for $Q=12$.}
    \label{tab::coul_lin_comp}
\end{table}

Finally, a system of three identical non-relativistic bosons that interacts through two-body attractive and three-body repulsive Gaussian interactions is considered. The ground and first excited state energies for such a system are provided in \cite[p.87]{baye15} and in \cite{gatt11}, respectively by mean of a LMM and an HHE. These references use
\begin{align}
    &V_{12}(r) = V_{23}(r) = V_{13}(r) = - V_0 e^{-r_{12}^2/R^2}, && W(\rho) = W_0 e^{-\rho^2/\rho_0^2}. \label{eq::four_acc_test}
\end{align}
with $V_0=1.227$ K, $W_0=0.279$ K, $R=10.03$ a.u., $\rho_0=13.85$ a.u. and with masses of $0.0231048\ (\text{a.u.})^{-2}\text{K}^{-1}$. In \cite{gatt11,baye15}, this potential is used to describe Helium trimers. The energy and mean-value of the inter-particle distance provided by the OBE, by the LMM and by the HHE are presented in Table \ref{tab::gauss_att_rep_comp}. The ground state energy upper bound given by the OBE is in agreement with both other results. The mean-values were not computed in \cite{gatt11} but both OBE and LMM evaluations are compatible. In addition to the ground state, an excited state is mentioned in \cite{baye15,gatt11}. This state is so weakly bound that the OBE cannot provide any approximation for it, even up to $30$ quanta.
\begin{table}
    \centering
    \begin{tabular}{l|l l l}
    $\ket{1;1;0^+}$ & \hspace{1mm} OBE \hspace{10mm} & LMM \hspace{10mm} & HHE \\
    \hline \hline
    $E$ (K) \hspace{1mm} & \hspace{1mm} $-0.1263$ & $-0.1264$ & $-0.1264$ \\
    $\braket{r}$ (a.u.) \hspace{1mm} & \hspace{1mm} $17.4010$ & $17.4707$ & N.A.
    \end{tabular}
    \caption{Ground state energies, $E$, and mean values for the inter-particle distance, $\braket{r}$, are given for a system of three identical bosons interacting with potentials \eqref{eq::four_acc_test}. Results from the OBE, from the literature (LLM and HHE) are compared. For HHE, mean-value is not provided in the literature. The chosen maximal number of quanta in the basis is $Q_{\text{max}}=24$ with $a$ computed for $Q=12$.}
    \label{tab::gauss_att_rep_comp}
\end{table}


\subsection{Complexity Tests}
\label{ssec:CTest}

To demonstrate that hyperspherical coefficients indeed accelerate the evaluation of matrix elements, a comparison in runtime is conducted with the naive method introduced in Section \ref{sec::3BMatEl}. This comparison involves a three-boson system with non-relativistic kinematics, unit masses, and a single attractive linear three-body potential, as defined below,
\begin{align}
    &V_{12}(r)=V_{23}(r)=V_{13}(r)=0, && W(\rho)=\frac{1}{2}\rho. \label{eq::comp_test_1}
\end{align}
Ground-state evaluations, without optimisation on the variational parameter, are executed for different basis lengths (using Python3 for programming). The utilisation of hyperspherical coefficients appears to reduce the runtime by approximately two orders of magnitude. The evaluation time as a function of the number of states in the basis roughly follows a power law with an exponent of $3.2$. Knowing the duration for a specific basis size, this indicative fit allows to infer the time required for evaluations with an increased number of quanta. However, let us remind that hyperspherical coefficients have been precomputed and stored in advance, requiring only their extraction during each run. This is possible as hyperspherical coefficients are independent of the specific three-body interaction considered. The evaluation of hyperspherical coefficients with formula \eqref{eq::hyperCoef_formula4} seems to roughly follow a power law with an exponent of $2.0$.

Hyperspherical coefficients can also be used to handle systems containing two identical particles and a distinct one. However, this necessitates forfeiting one of the two nonlinear variational parameters, consequently diminishing the accuracy of the resultant upper bounds. Despite this trade-off, the potential time savings during evaluations may enable the exploration of larger basis sizes. Thus, it remains uncertain whether the use of hyperspherical coefficients will enhance overall accuracy. To address this query, a test is conducted on the ground state of a system involving two bosons of mass $1$ and one particle of mass $10$, interacting through the three-body potential \eqref{eq::comp_test_1}. Allocating $500$ seconds for both methods, their respective accuracies are compared. Because the optimisation process duration depends on the inserted seed for $a$ and $b$, in the test, the user has been assumed to possess the first digit of the optimised values of $a$ and $b$. The conclusion is that hyperspherical coefficients deliver a rapid and precise upper bound for the ground state while two-dimensional integrals struggles to achieve a commensurate level of accuracy even after $500$ seconds. It turns out that, using the two-parameters method, the optimised value of $b$ aligns in magnitude with $\sqrt{3}a/2$ for the chosen system. This feature can explain the low loss of accuracy upon setting from the begining $b=\sqrt{3}a/2$.


\section{Possible generalisations}
\label{sec:gene}

The OBE method, in its most general form, extends beyond systems with merely three particles. While the hyperspherical coefficient method has been developed for systems of three identical particles, this approach readily accommodates handling three-body forces within systems containing a greater number of identical particles. In these instances, the harmonic oscillator eigenstates are constructed by sequentially coupling the $i^{\text{th}}$ Jacobi variable to the preceding $i-1$ variables,
\begin{equation}
    \Phi^{l_{12}l_{123}...L}_{n_1l_1...n_{N-1}l_{N-1}} \left(\bm{x_1},...,\bm{x_{N-1}}\right)=\left[...\left[\left[\phi_{n_1l_1}(\bm{x_1})\phi_{n_2l_2}(\bm{x_2})\right]_{l_{12}}\phi_{n_3l_3}(\bm{x_3})\right]_{l_{123}}...\right]_L
\end{equation}
where $\bm{x_i}$ represents the $i^{\text{th}}$ Jacobi coordinates within the $N$-body system. Consequently, any $N$-particle oscillator eigenstate encompasses a three-particle oscillator eigenstate. It allows to evaluate the matrix elements of the three-body interaction $W(\rho_{123})$ (with $\rho_{ijk}^2=r_{ij}^2+r_{ik}^2+r_{jk}^2$) can be evaluated using hyperspherical coefficients,
\begin{equation}
    \bal \bra{\Phi^{l_{12}'l_{123}'...L'}_{n_1'l_1'...n_{N-1}'l_{N-1}'} } W(\rho_{123}) \ket{\Phi^{l_{12}l_{123}...L}_{n_1l_1...n_{N-1}l_{N-1}}} = \delta_{n_3'n_3}&\delta_{l_3'l_3}...\delta_{n_{N-1}'n_{N-1}}\delta_{l_{N-1}'l_{N-1}}\delta_{l_{123}'l_{123}}...\delta_{L'L}\\ &\bra{\Phi^{l_{12}'}_{n_1'l_1'n_2'l_2'}(\bm{x_1},\bm{x_2})} W(\rho_{123}) \ket{\Phi^{l_{12}}_{n_1l_1n_2l_2}(\bm{x_1},\bm{x_2})}.
    \eal\label{eq::matEl_Npart}
\end{equation}
Furthermore, thanks to the symmetry properties of the wave-function for $N$ identical particles, all matrix elements associated with other triplets of particles are equal to those associated with particle $1$, $2$ and $3$,
\begin{equation}
    \bal
    \forall\ i,j,k\in\{1,...,N\},\ \bra{\Phi^{l_{12}'l_{123}'...L'}_{n_1'l_1'...n_{N-1}'l_{N-1}'}}& W(\rho_{ijk})\ket{\Phi^{l_{12}l_{123}...L}_{n_1l_1...n_{N-1}l_{N-1}}}\\
    &=\bra{\Phi^{l_{12}'l_{123}'...L'}_{n_1'l_1'...n_{N-1}'l_{N-1}'} } W(\rho_{123}) \ket{\Phi^{l_{12}l_{123}...L}_{n_1l_1...n_{N-1}l_{N-1}}}.
    \eal
\end{equation}
This implies that all the necessary matrix element are derived without the need for new formula.


\section{Conclusion}
\label{sec:conclu}

The OBE is a versatile and efficient approach for obtaining approximate eigenenergies and eigenfunctions in three-body systems. One of the key advantage of the OBE is its ability to easily handle both non- and semi-relativistic kinematics. While not explicitly presented in this paper, the formulas derived herein can be generalized to arbitrary kinematics. Such Hamiltonians are encountered, for example, in atomic physics \cite{arie92}, hadronic physics \cite{szcz96} and in quantum mechanics with a minimal length \cite{brau99}. The OBE also allows to easily provide symmetries, angular momentum and parity quantum numbers to the obtained approximation. In this study, the OBE, originally designed for two-body interactions, has been successfully extended to handle a specific class of three-body forces.

For systems of three identical particles, efficient computation of corresponding matrix elements is achieved by decomposing the trial states into harmonic oscillator eigenstates within hyperspherical coordinates. The coefficients of this decomposition are evaluated through an algebraic expression. This strategy enables matrix element evaluations at a rate two orders of magnitude faster than a naive strategy requiring two-dimensional numerical integrals. The extension of the OBE has been validated by comparing the approximations for the lowest eigenenergies and for the mean-values of interparticle distances to results obtained using the HHE and the LMM. The OBE with three-body forces have already been used in \cite{cimi23} to assess the accuracy of the envelope theory,  another approximation method in quantum mechanics.

In the case of systems comprising two identical particles and one distinct particle, the aforementioned decomposition remains applicable. However, it requires to give up one of the two non-linear variationnal parameters. An efficiency test on the ground state of a standard system demonstrated that the loss of accuracy due to this constraint can be compensated by the ability to reach higher basis sizes. Let us mention that the constraint $b=\sqrt{3}a/2$ reduces the OBE to its original form \cite{sema94,flec88}.

Because the implemented strategy fails to manage systems of three different particles, these were not investigated in the current study. Nevertheless, the method turned out to be easily generalisable to handle three-body forces in systems of $N$ identical bodies.


\section*{Acknowledgements} 

C.C. would like to thank the Fonds de la Recherche Scientifique - FNRS for the financial support. S.Y.K. would thank the University of Mons for her Initiation Research Grant. The authors thank J.Dohet-Eraly for the provided results as well as for his comments on this work and thank C.Semay and L.Cimino for the careful reading of the manuscript.


\appendix

\renewcommand{\thesection}{Appendix \Alph{section}}

\renewcommand{\theequation}{\Alph{section}.\arabic{equation}}

\section{Evaluation of the Hyperspherical coefficients}
\label{sec::hyperCoef}

The hyperspherical coefficients introduced in Section \ref{sec:theory} are defined in equation \eqref{eq::hyperCoef_def} as the overlap integral of two $\Phi^L_{n_xl_xn_yl_y}(\bm{x},\bm{y})$ and $\Phi_{NK}^{l_xl_y L}(\rho,\Omega)$ functions. This Appendix is dedicated to development of a closed formula that allows for an efficient evaluation of these coefficients. First, the six-dimensional volume element can be turned into hyperspherical coordinates \cite{das16},
\begin{equation}
    \diff^3x\, \diff^3y = \rho^5\sin(\theta_x)\sin(\theta_y)\sin^2(\alpha)\cos^2(\alpha)\,\diff\rho\,\diff\theta_x\,\diff\varphi_x\,\diff\theta_y\,\diff\varphi_y\,\diff\alpha. \label{eq::hyper_volEl}
\end{equation}
Using properties of spherical harmonics and Clebsh-Gordan coefficients \cite[sections 5.1.4 and 8.1.1]{vars88}, this integral on $\Phi$ and $\Psi$ functions reduces to an integral on $R$ and $\mathcal{R}$ functions,
\begin{equation}
    \bal
    \braket{n_xn_y|NK}_{l_xl_y} = \iint \rho^5 R_{n_xl_x}(\rho\sin\alpha)&R_{n_yl_y}(\rho\cos\alpha)\mathcal{R}_{NK}(\rho) N_K^{l_xl_y} \\ &\sin^{l_x+2}(\alpha)\cos^{l_y+2}(\alpha) P^{\left(l_x+\frac{1}{2},l_y+\frac{1}{2}\right)}_{n}(\cos 2\alpha)\,\diff\rho\, \diff\alpha.
    \label{eq::hyperCoef_formula1}
    \eal
\end{equation}
Let us remind that $n$ is a notation shortcut for $(K-l_x-l_y)/2$. Substituting $R$ and $\mathcal{R}$ functions by their definitions \eqref{eq::eig_1OH} and \eqref{eq::hyperEig1}, the integral becomes
\begin{equation}
    \bal
    \braket{n_xn_y|NK}_{l_xl_y} = C \int \diff\rho\,\rho^{5+l_x+l_y+K} e^{-\rho^2} L_N^{K+2}&(\rho^2) \int \diff\alpha\, \sin^{2l_x+2}(\alpha)\cos^{2l_y+2}(\alpha) \\
    & P^{\left(l_x+\frac{1}{2},l_y+\frac{1}{2}\right)}_{n}(\cos 2\alpha)
    L^{l_x+1/2}_{n_x}(\rho^2\sin^2\alpha)
    L^{l_y+1/2}_{n_y}(\rho^2\cos^2\alpha)
    \label{eq::hyperCoef_formula2}
    \eal
\end{equation}
where $C$ is a constant factor that gathers all the normalisation coefficients,
\begin{equation}
    C = N_K^{l_xl_y}\left(\frac{8(n_x!)(n_y!)(N!)}{\Gamma(n_x+l_x+3/2) \Gamma(n_y+l_y+3/2)\Gamma(K+N+3)}\right)^{1/2}. \label{eq::C_def}
\end{equation}
To get rid of the generalized Laguerre and Jacobi polynomials, one can replace them by their polynomial expressions \cite[formulas 22.3.1 and 22.3.9]{abra64},
\begin{align}
    L^{a}_{n} (x) &= \sum_{i=0}^{n} (-1)^i\binom{n + a}{n - i}\frac{x^i}{i!}, \label{eq::polyLag}\\
    P^{(a,b)}_{n} (x) &= 
    \sum_{i = 0}^{n} \binom{n + b}{n - i}\binom{n + a}{i}\left(\frac{x+1}{2}\right)^{i} \left(\frac{x-1}{2}\right)^{n-i}. \label{eq::polyJaco}
\end{align}
The symbol $\binom{x}{y}$ refers to the usual binomial coefficient defined in terms of gamma functions as
\begin{equation}
    \binom{x}{y}=\frac{\Gamma(x+1)}{\Gamma(y+1)\Gamma(x-y+1)}. \label{eq::binom_def}
\end{equation}
Once these replacements are performed, the hyperspherical coefficients are expressed as a the result of four sums,
\begin{equation}
    \bal
    \braket{n_xn_y|NK}_{l_xl_y} = C \sum_{s=0}^N& \frac{(-1)^s}{s!}\binom{N + K + 2}{N - s} 
    \sum_{m=0}^n (-1)^{n - m} \binom{n + l_y + \frac{1}{2}}{n-m}\binom{n + l_x + \frac{1}{2}}{m}\\
    &\sum_{i=0}^{n_x} \frac{(-1)^i}{i!}\binom{n_x + l_x + \frac{1}{2}}{n_x - i} 
    \sum_{j=0}^{n_y} \frac{(-1)^j}{j!}\binom{n_y + l_y + \frac{1}{2}}{n_y - j}\\
    &\phantom{\sum_{i=0}^{n_x}}\int\diff\rho\,\rho^{5+l_x+l_y+K+2i+2j+2s} e^{-\rho^2}\int \diff \alpha\, \left(\sin\alpha\right)^{2(l_x+1+i+n-m)}\left(\cos\alpha\right)^{2(l_y+1+j+m)}.
    \label{eq::hyperCoef_formula3}
    \eal
\end{equation}
The integrals on $\rho$ and $\alpha$ being decoupled, they can be evaluated and expressed in terms of gamma functions \cite[formulas 6.2.1, 6.2.2, 7.4.4 and 7.4.5]{abra64}, 
\begin{align}
    2\int_{0}^{\infty}\diff\rho\,\rho^{i} e^{-\rho^2} & = \Gamma\left(\frac{i+1}{2}\right),\label{eq::power_int}\\
    2\int_{0}^{\pi/2} \diff \alpha\, \left(\sin\alpha\right)^{i}\left(\cos\alpha\right)^{j} &= \frac{\Gamma\left(\frac{i+1}{2}\right)\Gamma\left(\frac{j+1}{2}\right)}{\Gamma\left(\frac{i+j}{2}+1\right)}.\label{eq::sincos_int}
\end{align}
This leads to the final expression,
\begin{equation}
    \bal
    \braket{n_xn_y|NK}_{l_xl_y} = C \sum_{s=0}^N& \frac{(-1)^s}{s!}\binom{N + K + 2}{N - s} 
    \sum_{m=0}^n (-1)^{n-m} \binom{n + l_y + \frac{1}{2}}{n-m}\binom{n + l_x + \frac{1}{2}}{m}\\
    &\sum_{i=0}^{n_x} \frac{(-1)^i}{i!}\binom{n_x + l_x + \frac{1}{2}}{n_x - i} 
    \sum_{j=0}^{n_y} \frac{(-1)^j}{j!}\binom{n_y + l_y + \frac{1}{2}}{n_y - j}\\
    &\phantom{\sum_{i=0}^{n_x}}
    \frac{\Gamma\left(\frac{l_x+l_y+K}{2}+3+i+j+s\right)\Gamma\left(l_x+i+n-m+\frac{3}{2}\right)\Gamma\left(l_y+j+m+\frac{3}{2}\right)}{4\,\Gamma\left(l_x+l_y+n+j+i+3\right)}.
    \label{eq::hyperCoef_formula4}
    \eal
\end{equation}
This formula enables to compute accurately all the hyperspherical coefficients. It is consistent with the formula for $\braket{n_xn_y|0K}_{l_xl_y}$ given in \cite{rayn70}. As the value of $2n_x+l_x+2n_y+l_y$ increases, the number of coefficients to compute as well as the time needed to evaluate the four sums also increase. Nevertheless, since the hyperspherical coefficients does not depend of any parameters, they do not have to be recomputed every time. A single evaluation beforehand and the storage of the result in a data file allows to retrieve the coefficients efficiently. Using the Mathematica software, all the hyperspherical coefficients until $2n_x+l_x+2n_y+l_y=40$  have been computed in a few hours on a regular laptop computer. Both $\Phi^L_{n_xl_xn_yl_y}$ and $\Psi_{NK}^{l_xl_yL}$ sets of states from relation \eqref{eq::hyperExp} being orthonormalised, hyperspherical coefficients satisfy the following probability conservation relation,
\begin{equation}
    \sum_{N\,K} \braket{n_xn_y|NK}_{l_xl_y}^2 = 1 \label{eq::hyperCoef_check}
\end{equation}
with the constraint $2N+K=2n_x+l_x+2n_y+l_y$ in the summations. This relation provides a useful consistency check to estimate the accuracy of the computed coefficients. 


\section{Analytical formulas for Talmi's integrals}
\label{sec::TalmiInt}

This Appendix provides analytical expressions of Talmi's integrals \eqref{eq::TalmiIntegral2} associated with the few $\mathcal{O}$ functions that are used in this paper. For power potentials (including the linear and Coulomb potentials used in Section \ref{ssec:ATest}), these integrals are easily expressed in terms of Gamma functions,
\begin{equation}
    I_p\left(\mathcal{O}(x) = \alpha x^\beta,a\right) = \alpha\,a^\beta\,\frac{\Gamma(p+3/2+\beta/2)}{\Gamma(p+3/2)}.
\end{equation}
Gaussian potentials (also employed in Section \ref{ssec:ATest}) admit analytical expressions as well,
\begin{equation}
    I_p\left(\mathcal{O}(x)=\alpha e^{-\beta\,x^2},a\right)= \alpha\,(1 + a^2 \beta)^{-3/2-p}.
\end{equation}
The treatment of semi-relativistic kinetic energies in Section \ref{sssec::RelKin} relies on some Talmi's integrals that also possess analytical expressions \cite{silv20},
\begin{equation}
    I_p\left(\mathcal{O}(x)=\sqrt{x^2 + \alpha},a\right)= a\left(\frac{\alpha}{a^2}\right)^{p+2}
    U\left(p+\frac{3}{2},p+3,\frac{\alpha}{a^2}\right),
\end{equation}
where $U$ denotes confluent hypergeometric functions of the second kind \cite{abra64}. 


\bibliographystyle{elsarticle-num-names.bst}
\bibliography{ref}

\begin{thebibliography}{31}
\expandafter\ifx\csname natexlab\endcsname\relax\def\natexlab#1{#1}\fi
\providecommand{\url}[1]{\texttt{#1}}
\providecommand{\href}[2]{#2}
\providecommand{\path}[1]{#1}
\providecommand{\DOIprefix}{doi:}
\providecommand{\ArXivprefix}{arXiv:}
\providecommand{\URLprefix}{URL: }
\providecommand{\Pubmedprefix}{pmid:}
\providecommand{\doi}[1]{\href{http://dx.doi.org/#1}{\path{#1}}}
\providecommand{\Pubmed}[1]{\href{pmid:#1}{\path{#1}}}
\providecommand{\bibinfo}[2]{#2}
\ifx\xfnm\relax \def\xfnm[#1]{\unskip,\space#1}\fi
\bibitem[{J.K.L.MacDonald(1933)}]{macd33}
\bibinfo{author}{J.K.L.MacDonald},
\newblock \bibinfo{title}{Successive approximations by the rayleigh-ritz
  variation method},
\newblock \bibinfo{journal}{Phys.Rev.} \bibinfo{volume}{43}
  (\bibinfo{year}{1933}) \bibinfo{pages}{830}.
\bibitem[{Silvestre-Brac et~al.(2020)Silvestre-Brac, Bonnaz, Semay, and
  Brau}]{silv20}
\bibinfo{author}{B.~Silvestre-Brac}, \bibinfo{author}{R.~Bonnaz},
  \bibinfo{author}{C.~Semay}, \bibinfo{author}{F.~Brau},
  \bibinfo{title}{Quantum three body problems using harmonic oscillator bases
  with different sizes}, \bibinfo{year}{2020}.
  \href{http://arxiv.org/abs/2003.11028}{{\tt arXiv:2003.11028}},
  \bibinfo{note}{internal report ISN 0066}.
\bibitem[{S.Fleck et~al.(1988)S.Fleck, B.Silvestre-Brac, and
  J.M.Richard}]{flec88}
\bibinfo{author}{S.Fleck}, \bibinfo{author}{B.Silvestre-Brac},
  \bibinfo{author}{J.M.Richard},
\newblock \bibinfo{title}{Search for diquark clustering in baryons},
\newblock \bibinfo{journal}{Phys.Rev.D} \bibinfo{volume}{38}
  (\bibinfo{year}{1988}) \bibinfo{pages}{1519}.
\bibitem[{C.Semay and B.Silvestre-Brac(1994)}]{sema94}
\bibinfo{author}{C.Semay}, \bibinfo{author}{B.Silvestre-Brac},
\newblock \bibinfo{title}{Diquonia and potential models},
\newblock \bibinfo{journal}{Z.Phys.C} \bibinfo{volume}{61}
  (\bibinfo{year}{1994}) \bibinfo{pages}{271}.
\bibitem[{C.Semay et~al.(2005)C.Semay, F.Brau, and B.Silvestre-Brac}]{sema05}
\bibinfo{author}{C.Semay}, \bibinfo{author}{F.Brau},
  \bibinfo{author}{B.Silvestre-Brac},
\newblock \bibinfo{title}{Pentaquarks $uudd\bar s$ with one color sextet
  diquark},
\newblock \bibinfo{journal}{Phys.Rev.Lett.} \bibinfo{volume}{94}
  (\bibinfo{year}{2005}) \bibinfo{pages}{062001}.
\bibitem[{E.Klempt and B.C.Metsch(2012)}]{klem12}
\bibinfo{author}{E.Klempt}, \bibinfo{author}{B.C.Metsch},
\newblock \bibinfo{title}{Multiplet classification of light-quark baryons},
\newblock \bibinfo{journal}{Eur.Phys.J.A} \bibinfo{volume}{48}
  (\bibinfo{year}{2012}) \bibinfo{pages}{127}.
\bibitem[{S.Noh et~al.(2021)S.Noh, W.Park, and S.H.Lee}]{noh21}
\bibinfo{author}{S.Noh}, \bibinfo{author}{W.Park}, \bibinfo{author}{S.H.Lee},
\newblock \bibinfo{title}{Doubly heavy tetraquarks, $qq'\bar q \bar q'$, in a
  nonrelativistic quark model with a complete set of harmonic oscillator
  bases},
\newblock \bibinfo{journal}{Phys.Rev.D} \bibinfo{volume}{103}
  (\bibinfo{year}{2021}) \bibinfo{pages}{114009}.
\bibitem[{M.Gattobigio et~al.(2011)M.Gattobigio, A.Kievsky, and
  M.Viviani}]{gatt11}
\bibinfo{author}{M.Gattobigio}, \bibinfo{author}{A.Kievsky},
  \bibinfo{author}{M.Viviani},
\newblock \bibinfo{title}{Spectra of helium clusters with up to six atoms using
  soft-core potentials.},
\newblock \bibinfo{journal}{Phys.Rev.A} \bibinfo{volume}{84}
  (\bibinfo{year}{2011}) \bibinfo{pages}{052503}.
\bibitem[{F.Buisseret et~al.(2022)F.Buisseret, Willemyns, and C.Semay}]{buis22}
\bibinfo{author}{F.Buisseret}, \bibinfo{author}{C.~Willemyns},
  \bibinfo{author}{C.Semay},
\newblock \bibinfo{title}{Many-quark interactions: Large-$n$ scaling and
  contribution to baryon masses},
\newblock \bibinfo{journal}{Universe} \bibinfo{volume}{8}
  (\bibinfo{year}{2022}) \bibinfo{pages}{311}.
\bibitem[{B.Desplanques et~al.(1992)B.Desplanques, C.Gignoux, B.Silvestre-Brac,
  P.Gonz\'{a}lez, J.Navarro, and Noguera}]{desp92}
\bibinfo{author}{B.Desplanques}, \bibinfo{author}{C.Gignoux},
  \bibinfo{author}{B.Silvestre-Brac}, \bibinfo{author}{P.Gonz\'{a}lez},
  \bibinfo{author}{J.Navarro}, \bibinfo{author}{S.~Noguera},
\newblock \bibinfo{title}{The baryonic spectrum in a constituent quark model
  including a three-body force},
\newblock \bibinfo{journal}{Z.Phys.A} \bibinfo{volume}{343}
  (\bibinfo{year}{1992}) \bibinfo{pages}{331}.
\bibitem[{S.Pepin and F.Stancu(2002)}]{pepi92}
\bibinfo{author}{S.Pepin}, \bibinfo{author}{F.Stancu},
\newblock \bibinfo{title}{Three-body confinement force in hadron spectroscopy},
\newblock \bibinfo{journal}{Phys.Rev.D} \bibinfo{volume}{65}
  (\bibinfo{year}{2002}) \bibinfo{pages}{054032}.
\bibitem[{V.Dmitra\v{s}inovi\`{c}(2001)}]{dmit01}
\bibinfo{author}{V.Dmitra\v{s}inovi\`{c}},
\newblock \bibinfo{title}{Cubic casimir operator of $su_c(3)$ and confinement
  in the nonrelativistic quark model},
\newblock \bibinfo{journal}{Phys.Lett.B} \bibinfo{volume}{499}
  (\bibinfo{year}{2001}) \bibinfo{pages}{135}.
\bibitem[{M.Ferraris et~al.(1995)M.Ferraris, M.M.Giannini, M.Pizzo,
  E.Santopinto, and L.Tiator}]{ferr95}
\bibinfo{author}{M.Ferraris}, \bibinfo{author}{M.M.Giannini},
  \bibinfo{author}{M.Pizzo}, \bibinfo{author}{E.Santopinto},
  \bibinfo{author}{L.Tiator},
\newblock \bibinfo{title}{A three-body force model for the baryon spectrum},
\newblock \bibinfo{journal}{Phys.Lett.B} \bibinfo{volume}{364}
  (\bibinfo{year}{1995}) \bibinfo{pages}{231}.
\bibitem[{S.Ishikawa(2017)}]{ishi17}
\bibinfo{author}{S.Ishikawa},
\newblock \bibinfo{title}{Three-body potentials in $\alpha$-particle model of
  light nuclei},
\newblock \bibinfo{journal}{Few-Body Syst.} \bibinfo{volume}{58}
  (\bibinfo{year}{2017}) \bibinfo{pages}{37}.
\bibitem[{T.K.Das(2016)}]{das16}
\bibinfo{author}{T.K.Das}, \bibinfo{title}{Hyperspherical Harmonics Expansion
  Techniques}, \bibinfo{publisher}{Springer}, \bibinfo{address}{New Delhi},
  \bibinfo{year}{2016}.
\bibitem[{J.Raynal and J.Revai(1970)}]{rayn70}
\bibinfo{author}{J.Raynal}, \bibinfo{author}{J.Revai},
\newblock \bibinfo{title}{Transformation coefficients in the hyperspherical
  approach to the three-body problem},
\newblock \bibinfo{journal}{Nuovo Cimento A (1965-1970)} \bibinfo{volume}{68}
  (\bibinfo{year}{1970}) \bibinfo{pages}{612}.
\bibitem[{V.K.Khersonskii et~al.(1988)V.K.Khersonskii, A.N.Moskalev, and
  D.A.Varshalovich}]{vars88}
\bibinfo{author}{V.K.Khersonskii}, \bibinfo{author}{A.N.Moskalev},
  \bibinfo{author}{D.A.Varshalovich}, \bibinfo{title}{Quantum Theory Of Angular
  Momentum}, \bibinfo{publisher}{World Scientific Publishing Co.},
  \bibinfo{address}{Singapore}, \bibinfo{year}{1988}.
\bibitem[{M.Abramowitz and I.A.Stegun(1964)}]{abra64}
\bibinfo{author}{M.Abramowitz}, \bibinfo{author}{I.A.Stegun},
  \bibinfo{title}{Handbook of Mathematical Functions with Formulas, Graphs, and
  Mathematical Tables}, \bibinfo{publisher}{Dover Publications, Inc.},
  \bibinfo{address}{New York}, \bibinfo{year}{1964}.
\bibitem[{T.A.Brody et~al.(1960)T.A.Brody, G.Jacob, and Moshinsky}]{brod60}
\bibinfo{author}{T.A.Brody}, \bibinfo{author}{G.Jacob},
  \bibinfo{author}{M.~Moshinsky},
\newblock \bibinfo{title}{Matrix elements in nuclear shell theory},
\newblock \bibinfo{journal}{Nucl.Phys.} \bibinfo{volume}{17}
  (\bibinfo{year}{1960}) \bibinfo{pages}{16}.
\bibitem[{M.Moshinsky(1969)}]{mosh69}
\bibinfo{author}{M.Moshinsky}, \bibinfo{title}{The Harmonic Oscillator in
  Modern Physics: from Atoms to Quarks}, \bibinfo{publisher}{Gordon and Breach,
  Science Publishers, Inc.}, \bibinfo{address}{New York}, \bibinfo{year}{1969}.
\bibitem[{B.Silvestre-Brac(1985)}]{silv85}
\bibinfo{author}{B.Silvestre-Brac},
\newblock \bibinfo{title}{The cluster model and the generalized brody-moshinsky
  coefficients},
\newblock \bibinfo{journal}{J.Physique} \bibinfo{volume}{46}
  (\bibinfo{year}{1985}) \bibinfo{pages}{1087}.
\bibitem[{P.Nunberg et~al.(1977)P.Nunberg, D.Prosperi, and E.Pace}]{nunb77}
\bibinfo{author}{P.Nunberg}, \bibinfo{author}{D.Prosperi},
  \bibinfo{author}{E.Pace},
\newblock \bibinfo{title}{An application of a new harmonic-oscillator basis to
  the calculation of trinucleon ground-state observables},
\newblock \bibinfo{journal}{Nucl.Phys.A} \bibinfo{volume}{285}
  (\bibinfo{year}{1977}) \bibinfo{pages}{58}.
\bibitem[{A.Kievsky et~al.(2008)A.Kievsky, S.Rosati, M.Viviani, L.E.Marcucci,
  and L.Girlanda}]{kiev08}
\bibinfo{author}{A.Kievsky}, \bibinfo{author}{S.Rosati},
  \bibinfo{author}{M.Viviani}, \bibinfo{author}{L.E.Marcucci},
  \bibinfo{author}{L.Girlanda},
\newblock \bibinfo{title}{A high-precision variational approach to three- and
  four-nucleon bound and zero-energy scattering states},
\newblock \bibinfo{journal}{J.Phys.G: Nucl.Part.Phys.} \bibinfo{volume}{35}
  (\bibinfo{year}{2008}) \bibinfo{pages}{063101}.
\bibitem[{L.E.Marcucci et~al.(2020)L.E.Marcucci, J.Dohet-Eraly, L.Girlanda,
  A.Gnech, A.Kievsky, and M.Viviani}]{marc20}
\bibinfo{author}{L.E.Marcucci}, \bibinfo{author}{J.Dohet-Eraly},
  \bibinfo{author}{L.Girlanda}, \bibinfo{author}{A.Gnech},
  \bibinfo{author}{A.Kievsky}, \bibinfo{author}{M.Viviani},
\newblock \bibinfo{title}{The hyperspherical harmonics method: A tool for
  testing and improving nuclear interaction models},
\newblock \bibinfo{journal}{Front.Phys.} \bibinfo{volume}{8}
  (\bibinfo{year}{2020}) \bibinfo{pages}{69}.
\bibitem[{D.Baye(2015)}]{baye15}
\bibinfo{author}{D.Baye},
\newblock \bibinfo{title}{The lagrange-mesh method},
\newblock \bibinfo{journal}{Phys.Rep.} \bibinfo{volume}{565}
  (\bibinfo{year}{2015}) \bibinfo{pages}{1}.
\bibitem[{S.Capstick and N.Isgur(1986)}]{caps86}
\bibinfo{author}{S.Capstick}, \bibinfo{author}{N.Isgur},
\newblock \bibinfo{title}{Baryons in a relativized quark model with
  chromodynamics},
\newblock \bibinfo{journal}{Phys.Rev.D} \bibinfo{volume}{34}
  (\bibinfo{year}{1986}) \bibinfo{pages}{2809}.
\bibitem[{J.Dohet-Eraly(2024)}]{PC}
\bibinfo{author}{J.Dohet-Eraly}, \bibinfo{year}{2024}. \bibinfo{note}{Private
  communication}.
\bibitem[{Altschul et~al.(1992)Altschul, A.Fraenkel, and E.Finkman}]{arie92}
\bibinfo{author}{V.~Altschul}, \bibinfo{author}{A.Fraenkel},
  \bibinfo{author}{E.Finkman},
\newblock \bibinfo{title}{Effects of band nonparabolicity on twodimensionnal
  electron gas},
\newblock \bibinfo{journal}{J.Appl.Phys.} \bibinfo{volume}{71}
  (\bibinfo{year}{1992}) \bibinfo{pages}{4382}.
\bibitem[{A.Szczepaniak et~al.(1996)A.Szczepaniak, E.S.Swanson, C-R.Ji, and
  S.R.Cotanch}]{szcz96}
\bibinfo{author}{A.Szczepaniak}, \bibinfo{author}{E.S.Swanson},
  \bibinfo{author}{C-R.Ji}, \bibinfo{author}{S.R.Cotanch},
\newblock \bibinfo{title}{Glueball spectroscopy in a relativistic many-body
  approach to hadronic structure},
\newblock \bibinfo{journal}{Phys.Rev.Lett.} \bibinfo{volume}{76}
  (\bibinfo{year}{1996}) \bibinfo{pages}{2011}.
\bibitem[{F.Brau(1999)}]{brau99}
\bibinfo{author}{F.Brau},
\newblock \bibinfo{title}{Minimal length uncertainty relation and the hydrogen
  atom},
\newblock \bibinfo{journal}{J.Phys.A:Math.Gen.} \bibinfo{volume}{32}
  (\bibinfo{year}{1999}) \bibinfo{pages}{7691}.
\bibitem[{Cimino et~al.(2024)Cimino, Tourbez, Chevalier, Lacroix, and
  Semay}]{cimi23}
\bibinfo{author}{L.~Cimino}, \bibinfo{author}{C.~Tourbez},
  \bibinfo{author}{C.~Chevalier}, \bibinfo{author}{G.~Lacroix},
  \bibinfo{author}{C.~Semay},
\newblock \bibinfo{title}{Tests of the envelope theory for three-body forces},
\newblock \bibinfo{journal}{Few-Body Syst.} \bibinfo{volume}{65}
  (\bibinfo{year}{2024}) \bibinfo{pages}{20}.

\end{thebibliography}


\end{document}